\documentclass[aps,prd,preprint,tightenlines,superscriptaddress,showpacs,byrevtex]{revtex4}
\usepackage{graphicx}% Include figure files
\usepackage{dcolumn}% Align table columns on decimal point
\usepackage{bm}% bold math
\def\bz{{B^0}}
\def\bzb{{\overline{B}{}^0}}
\def\bp{{B^+}}

\def\kl{K_L^0}
\def\dE{{\Delta E}}
\def\mb{{M_{\rm bc}}}
\def\Dt{\Delta t}
\def\Dz{\Delta z}
\def\fol{f_{\rm ol}}
\def\fsig{f_{\rm sig}}
\newcommand{\sinbb}{{\sin2\phi_1}}

\newcommand{\fCP}{f_{CP}}

\newcommand{\ftag}{f_{\rm tag}}

\newcommand{\cala}{{\cal A}}
\newcommand{\cals}{{\cal S}}
\newcommand{\dm}{\Delta m_d}
\newcommand{\dmd}{\dm}
\def\taubz{{\tau_\bz}}
\def\taubp{{\tau_\bp}}
\def\ks{{K_S^0}}

\newcommand{\btosss}{b \to s\overline{s}s}
\newcommand*{\dwl}{\ensuremath{{\Delta w_l}}}
\newcommand*{\fq}{\ensuremath{q}}
\def\kz{{K^0}}
\def\kp{{K^+}}
\def\km{{K^-}}
\def\fzero{{f_0(980)}}
\def\fx{{f_X(1300)}}
\def\pip{{\pi^+}}
\def\pim{{\pi^-}}
\def\piz{{\pi^0}}
\def\kstarz{{K^{*0}}}
\def\kstarp{{K^{*+}}}

\def\kl{{K_L^0}}
\def\bbar{{\overline{B}}}
\def\ufs{{\Upsilon(4S)}}

\def\nsig{{N_{\rm sig}}}

\def\nsigmc{{N_{\rm sig}^{\rm MC}}}
\def\nbkg{{N_{\rm bkg}}}

\def\jpsi{{J/\psi}}

\def\rhoz{{\rho^0}}

\newcommand{\dslnu}{D^{*-}\ell^+\nu}
\newcommand{\bzdslnu}{\bz \to \dslnu}

\def\sperp{{S_{\perp}}}
\def\lsig{{\cal L}_{\rm sig}}
\def\lbkg{{\cal L}_{\rm bkg}}
\def\rsigbkg{{\cal R}_{\rm s/b}}
\def\calf{{\cal F}}
\def\rkpi{{\cal R}_{K/\pi}}

\def\mgg{M_{\gamma\gamma}}
\def\ppizcms{p_\piz^{\rm cms}}

\def\pbstar{p_B^{\rm cms}}

\newcommand*{\eeff}{\ensuremath{\epsilon_\textrm{eff}}}

\def\lnsig{{\cal L}_{N_{\rm sig}}}
\def\lzero{{\cal L}_0}

\def\efftot{{0.30\pm 0.01}}
\def\efftotdsone{{0.287 \pm 0.005}}

\def\sinbbWA{+0.726}
\def\sinbbERR{0.037}
\def\sinbbWAResult{\sinbbWA\pm\sinbbERR}

\def\SphiksResPrv{-0.96\pm0.50^{+0.09}_{-0.11}}
\def\AphiksResPrv{-0.15\pm0.29\pm0.07}

   \def\Pphiks{0.63}     \def\Nsigphiks{139 \pm 14}

   \def\Pphikl{0.17}     \def\Nsigphikl{36 \pm 15}

  \def\Pkpkmks{0.56}    \def\Nsigkpkmks{398\pm 28}

   \def\PkspizH{0.54}     \def\NsigkspizH{168\pm 16}

   \def\PkspizL{0.15}     \def\NsigkspizL{79\pm 19}

  \def\Petapks{0.61}    \def\Nsigetapks{512 \pm 27}

 \def\Pomegaks{0.56}    \def\Nsigomegaks{31 \pm 7}

 \def\Pfzeroks{0.53}    \def\Nsigfzeroks{94 \pm 14}
\def\SphikzVal{+0.08}    \def\SphikzStat{0.33}    \def\SphikzSyst{0.09}
\def\AphikzVal{+0.08}    \def\AphikzStat{0.22}    \def\AphikzSyst{0.09}
\def\SphiksVal{0.02}    \def\SphiksStat{0.33}    %\def\SphiksSyst{x.xx}
\def\AphiksVal{0.07}    \def\AphiksStat{0.22}    %\def\AphiksSyst{x.xx}
\def\mSphiklVal{2.3}    \def\mSphiklStat{2.0}    %\def\mSphiklSyst{x.xx}
\def\AphiklVal{0.6}     \def\AphiklStat{1.2}     %\def\AphiklSyst{x.xx}
\def\SkpkmksVal{-0.49}   \def\SkpkmksStat{0.18}   \def\SkpkmksSyst{0.04}
\def\AkpkmksVal{-0.09}   \def\AkpkmksStat{0.12}   \def\AkpkmksSyst{0.07}

\def\SkspizVal{+0.32}    \def\SkspizStat{0.61}    \def\SkspizSyst{0.13}
\def\AkspizVal{-0.11}    \def\AkspizStat{0.20}    \def\AkspizSyst{0.09}
\def\SetapksVal{+0.65}   \def\SetapksStat{0.18}   \def\SetapksSyst{0.04}
\def\AetapksVal{-0.19}   \def\AetapksStat{0.11}   \def\AetapksSyst{0.05}
\def\SomegaksVal{+0.76}  \def\SomegaksStat{0.65}  \def\SomegaksSyst{^{+0.13}_{-0.16}}
\def\AomegaksVal{+0.27}  \def\AomegaksStat{0.48}  \def\AomegaksSyst{0.15}
\def\SfzeroksVal{+0.47}  \def\SfzeroksStat{0.41}  \def\SfzeroksSyst{0.08}
\def\AfzeroksVal{-0.39}  \def\AfzeroksStat{0.27}  \def\AfzeroksSyst{0.09}
\def\SbsqqVal{+0.40} \def\SbsqqErr{\pm0.13}

\def\SphikzResult{\SphikzVal\pm\SphikzStat\pm\SphikzSyst}

\def\AphikzResult{\AphikzVal\pm\AphikzStat\pm\AphikzSyst}

\def\SphiksResultStat{\SphiksVal\pm\SphiksStat}

\def\AphiksResultStat{\AphiksVal\pm\AphiksStat}

\def\mSphiklResultStat{\mSphiklVal\pm\mSphiklStat}

\def\AphiklResultStat{\AphiklVal\pm\AphiklStat}

\def\SkpkmksResult{\SkpkmksVal\pm\SkpkmksStat\pm\SkpkmksSyst}

\def\AkpkmksResult{\AkpkmksVal\pm\AkpkmksStat\pm\AkpkmksSyst}

\def\SkspizResult{\SkspizVal\pm\SkspizStat\pm\SkspizSyst}

\def\AkspizResult{\AkspizVal\pm\AkspizStat\pm\AkspizSyst}

\def\SetapksResult{\SetapksVal\pm\SetapksStat\pm\SetapksSyst}

\def\AetapksResult{\AetapksVal\pm\AetapksStat\pm\AetapksSyst}

\def\SomegaksResult{\SomegaksVal\pm\SomegaksStat\SomegaksSyst}

\def\AomegaksResult{\AomegaksVal\pm\AomegaksStat\pm\AomegaksSyst}

\def\SfzeroksResult{\SfzeroksVal\pm\SfzeroksStat\pm\SfzeroksSyst}

\def\AfzeroksResult{\AfzeroksVal\pm\AfzeroksStat\pm\AfzeroksSyst}

\def\SbsqqResult{\SbsqqVal\SbsqqErr}

\begin{document}

%\preprint{\vbox{
%\hbox{$Revision: 1.10 $}
%}}
\vspace*{-3\baselineskip}
\resizebox{!}{3cm}{\includegraphics{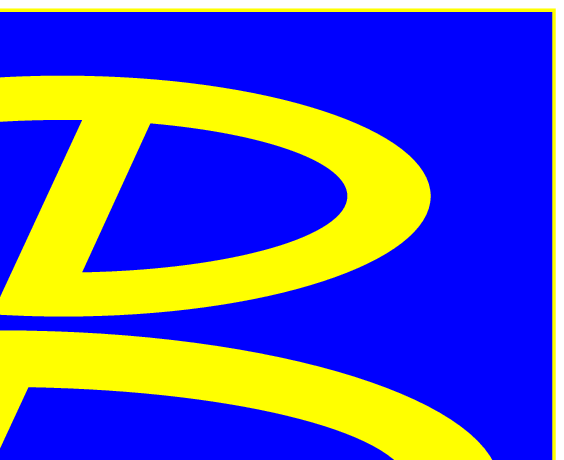}}
\vspace*{-3cm}
\begin{flushright}
Belle Preprint 2005-17\\
KEK   Preprint 2005-10
\end{flushright}
\vspace*{1cm}

%%%%%%%%%%%%%%%
% Paper Title
%%%%%%%%%%%%%%%
\title{\quad\\[0.5cm] \boldmath Time-Dependent $CP$-Violating 
       \\Asymmetries in $b \to s\overline{q}q$ Transitions}
%\title{\boldmath Time-Dependent $CP$-Violating 
%       Asymmetries in $b \to s\overline{q}q$ Transitions}

\date{\today}% It is always \today, today,

%%% Paper:    b -> s qbar q TCPV
%%% Journal:  Physical Review D
%%% Contacts: K. Hara (kohji@bmail.kek.jp)
%%%           K.-F. Chen (kfjack@hep1.phys.ntu.edu.tw)
%%%           M. Hazumi (masashi.hazumi@kek.jp)
%%% Non-responding authors or those who said NO are commented out.
%%% ====================================================================
%%% Click the RELOAD button on your web browser to see the updated file.
%%% ====================================================================
%%% Use \input{author} to insert this material into your latex file.
%%%%% Force institutions to appear in alphabetical order when typeset.
%%%\affiliation{Aomori University, Aomori}
\affiliation{Budker Institute of Nuclear Physics, Novosibirsk}
\affiliation{Chiba University, Chiba}
\affiliation{Chonnam National University, Kwangju}
\affiliation{University of Cincinnati, Cincinnati, Ohio 45221}
%%%\affiliation{University of Frankfurt, Frankfurt}
\affiliation{Gyeongsang National University, Chinju}
\affiliation{University of Hawaii, Honolulu, Hawaii 96822}
\affiliation{High Energy Accelerator Research Organization (KEK), Tsukuba}
\affiliation{Hiroshima Institute of Technology, Hiroshima}
\affiliation{Institute of High Energy Physics, Chinese Academy of Sciences, Beijing}
\affiliation{Institute of High Energy Physics, Vienna}
\affiliation{Institute for Theoretical and Experimental Physics, Moscow}
\affiliation{J. Stefan Institute, Ljubljana}
\affiliation{Kanagawa University, Yokohama}
\affiliation{Korea University, Seoul}
%%%\affiliation{Kyoto University, Kyoto}
\affiliation{Kyungpook National University, Taegu}
\affiliation{Swiss Federal Institute of Technology of Lausanne, EPFL, Lausanne}
\affiliation{University of Ljubljana, Ljubljana}
\affiliation{University of Maribor, Maribor}
\affiliation{University of Melbourne, Victoria}
\affiliation{Nagoya University, Nagoya}
\affiliation{Nara Women's University, Nara}
\affiliation{National Central University, Chung-li}
%%%\affiliation{National Kaohsiung Normal University, Kaohsiung}
\affiliation{National United University, Miao Li}
\affiliation{Department of Physics, National Taiwan University, Taipei}
\affiliation{H. Niewodniczanski Institute of Nuclear Physics, Krakow}
\affiliation{Nippon Dental University, Niigata}
\affiliation{Niigata University, Niigata}
\affiliation{Nova Gorica Polytechnic, Nova Gorica}
\affiliation{Osaka City University, Osaka}
\affiliation{Osaka University, Osaka}
\affiliation{Panjab University, Chandigarh}
\affiliation{Peking University, Beijing}
\affiliation{Princeton University, Princeton, New Jersey 08544}
%%%\affiliation{RIKEN BNL Research Center, Upton, New York 11973}
\affiliation{Saga University, Saga}
\affiliation{University of Science and Technology of China, Hefei}
\affiliation{Seoul National University, Seoul}
\affiliation{Sungkyunkwan University, Suwon}
\affiliation{University of Sydney, Sydney NSW}
\affiliation{Tata Institute of Fundamental Research, Bombay}
\affiliation{Toho University, Funabashi}
\affiliation{Tohoku Gakuin University, Tagajo}
\affiliation{Tohoku University, Sendai}
\affiliation{Department of Physics, University of Tokyo, Tokyo}
\affiliation{Tokyo Institute of Technology, Tokyo}
\affiliation{Tokyo Metropolitan University, Tokyo}
\affiliation{Tokyo University of Agriculture and Technology, Tokyo}
%%%\affiliation{Toyama National College of Maritime Technology, Toyama}
\affiliation{University of Tsukuba, Tsukuba}
%%%\affiliation{Utkal University, Bhubaneswer}
\affiliation{Virginia Polytechnic Institute and State University, Blacksburg, Virginia 24061}
\affiliation{Yonsei University, Seoul}
% 12 first authors
   \author{K.-F.~Chen}\affiliation{Department of Physics, National Taiwan University, Taipei} % Taiwan
   \author{F.~Fang}\affiliation{University of Hawaii, Honolulu, Hawaii 96822} % Hawaii
   \author{A.~Garmash}\affiliation{Princeton University, Princeton, New Jersey 08544} % Princeton
   \author{K.~Hara}\affiliation{High Energy Accelerator Research Organization (KEK), Tsukuba} % KEK
   \author{M.~Hazumi}\affiliation{High Energy Accelerator Research Organization (KEK), Tsukuba} % KEK
   \author{T.~Higuchi}\affiliation{High Energy Accelerator Research Organization (KEK), Tsukuba} % KEK
   \author{H.~Kakuno}\affiliation{Department of Physics, University of Tokyo, Tokyo} % Tokyo
   \author{A.~Kusaka}\affiliation{Department of Physics, University of Tokyo, Tokyo} % Tokyo
   \author{T.~Shibata}\affiliation{Niigata University, Niigata} % Niigata
   \author{O.~Tajima}\affiliation{High Energy Accelerator Research Organization (KEK), Tsukuba} % KEK
% ...
% Ljubljana
% \author{D.~Z\"urcher}\affiliation{Swiss Federal Institute of Technology of Lausanne, EPFL, Lausanne} % Lausanne
\collaboration{The Belle Collaboration}

\begin{abstract}
  We present new measurements of $CP$-violation parameters
  in 
  $\bz\to$
  $\phi\kz$, 
  $\kp\km\ks$,
  $\fzero\ks$,
  $\eta'\ks$, 
  $\omega\ks$, 
  and
  $\ks\piz$
  decays
  based on a sample of $275\times 10^6$ $B\bbar$ pairs
  collected at the $\ufs$ resonance with
  the Belle detector at the KEKB energy-asymmetric $e^+e^-$ collider.
  One neutral $B$ meson is fully reconstructed in
  one of the specified decay channels,
  and the flavor of the accompanying $B$ meson is identified from
  its decay products.
  $CP$-violation parameters for each of the decay 
  modes are obtained from the asymmetries in the distributions of
  the proper-time intervals between the two $B$ decays.
  The combined result for
  the $\bz\to\phi\kz$, $\kp\km\ks$, $\fzero\ks$, $\eta'\ks$,
  $\omega\ks$,  $\ks\piz$
  and previously reported $\ks\ks\ks$ decays
  differs from the SM expectation by 2.4 standard deviations.
\end{abstract}


\maketitle

%%%%%%%%%%%%%%%%%%%%%%%%%%%
% Main Part
%%%%%%%%%%%%%%%%%%%%%%%%%%%
%%%%%%%%%%%%%%%%%%%%%%%
\section{Introduction}
\label{sec:introduction}
%%%%%%%%%%%%%%%%%%%%%%%

%$b \to s\overline{q}q$
The flavor-changing $b \to s$ transition proceeds through loop penguin
diagrams.
%Such loop diagrams is expected to be sensitive to new physics 
%to be sensitive to new physics 
Such loop diagrams play an important role in testing the standard model (SM) 
and new physics because particles beyond the SM can contribute via
additional loop diagrams.
$CP$ violation in the $b \to s$ transition is especially sensitive to physics
at a very high-energy scale~\cite{Akeroyd:2004mj}.
Theoretical studies indicate that
%large deviations from standard model (SM) expectations
large deviations from the SM expectations
are allowed for time-dependent $CP$ asymmetries in
$\bz$ meson decays~\cite{bib:lucy}.
Experimental investigations have recently been launched
at the two $B$ factories, each of which has produced more than
$10^8$ $B\bbar$ pairs.
Our previous measurement of the $CP$-violating asymmetry
in $\bz\to \phi\ks$~\cite{bib:CC},
which is dominated by the $\btosss$ transition,
yielded a value that differs from the SM expectation 
by 3.5 standard deviations~\cite{Abe:2003yt}.
Measurements with a larger data sample are required to 
elucidate this difference. It is also essential to examine
additional modes that
may be sensitive to the same $b \to s$ penguin amplitude.
In this spirit, experimental results using the decay modes
$\bz\to\phi\kl$, 
$\kp\km\ks$, 
$\fzero\ks$,
$\eta'\ks$, 
and
$\ks\piz$ have already been reported~\cite{Abe:2003yt,bib:BaBar_sss}.

In the SM, $CP$ violation arises from a single irreducible phase, 
the Kobayashi-Maskawa (KM) phase~\cite{Kobayashi:1973fv},
in the weak-interaction quark-mixing matrix. In
particular, the SM predicts $CP$ asymmetries in the time-dependent
rates for $\bz$ and
$\bzb$ decays to a common $CP$ eigenstate $\fCP$~\cite{bib:sanda}. 
In the decay chain $\Upsilon(4S)\to \bz\bzb \to f_{CP}f_{\rm tag}$,
where one of the $B$ mesons decays at time $t_{CP}$ to a 
final state $f_{CP}$ 
and the other decays at time $t_{\rm tag}$ to a final state  
$f_{\rm tag}$ that distinguishes between $B^0$ and $\bzb$, 
the decay rate has a time dependence
given by
\begin{equation}
\label{eq:psig}
{\cal P}(\Delta{t}) = 
\frac{e^{-|\Delta{t}|/{\taubz}}}{4{\taubz}}
\biggl\{1 + \fq\cdot 
\Bigl[ \cals\sin(\dmd\Delta{t})
   + \cala\cos(\dmd\Delta{t})
\Bigr]
\biggr\}.
\end{equation}
Here $\cals$ and $\cala$ are $CP$-violation parameters,
$\taubz$ is the $B^0$ lifetime, $\dmd$ is the mass difference 
between the two $B^0$ mass
eigenstates, $\Delta{t}$ = $t_{CP}$ $-$ $t_{\rm tag}$, and
the $b$-flavor charge $\fq$ = +1 ($-1$) when the tagging $B$ meson
is a $B^0$ ($\bzb$).
To a good approximation,
the SM predicts $\cals = -\xi_f\sin 2\phi_1$, where $\xi_f = +1 (-1)$
corresponds to  $CP$-even (-odd) final states, and $\cala =0$
for both $b \to c\overline{c}s$ and 
$b \to s\overline{q}q$ transitions.
Recent measurements of time-dependent $CP$ asymmetries in
$\bz \to J/\psi \ks$ and related decay 
modes by Belle~\cite{bib:CP1_Belle,bib:BELLE-CONF-0436}
and BaBar~\cite{bib:CP1_BaBar}, 
 which are governed by the $b \to c\overline{c}s$ transition,
have already determined $\sinbb$ rather precisely;
the present world average value is 
$\sinbb = \sinbbWAResult$~\cite{bib:HFAG}.
This serves as a firm reference point for the SM.

%%%%%%%%%%%%%%%%%%%%%%%%%%%%%%%%%%%%%%%%%%%%%%%%
%\subsection{Luminosity and Experimental Apparatus}
%%%%%%%%%%%%%%%%%%%%%%%%%%%%%%%%%%%%%%%%%%%%%%%%
Belle's previous measurements
for $\bz \to \phi\ks$, $K^+K^-\ks$ and $\eta'\ks$
were based on a 140 fb$^{-1}$ data sample (DS-I)
containing $152\times 10^6$ $B\bbar$ pairs.
In this report, we describe improved measurements 
incorporating an additional 113 fb$^{-1}$
data sample that contains $123\times 10^6$ $B\bbar$
pairs (DS-II) for a total of $275\times 10^6$ $B\bbar$ pairs.

While $\phi\ks$ and $\eta'\ks$ final states are
$CP$ eigenstates with $\xi_f=-1$, 
the $K^+K^-\ks$ final state is in general
a mixture of both $\xi_f=+1$ and $-1$.
Excluding $K^+K^-$ pairs that are consistent 
with a $\phi \to K^+K^-$ decay from the $\bz \to K^+K^-\ks$ sample,
we find that the $K^+K^-\ks$ state is primarily $\xi_f=+1$;
a measurement of the $\xi_f=+1$ fraction 
using the isospin relation~\cite{Garmash:2003er}
with a 253~fb$^{-1}$ data sample gives
$f_{+}=0.83 \pm 0.10\text{(stat)}\pm0.04\text{(syst)}$,
which is consistent with the previous result~\cite{Abe:2003yt}.
The SM expectation for this mode is
$\cals = -(2f_{+}-1)\sin 2\phi_1$.
%In the following determination of $\cals$ and $\cala$,
%we fix $\xi_f=+1$ for this mode.

We include additional $\phi\ks$ and $\eta'\ks$
subdecay modes that were not used in the previous analysis.
We also describe new measurements of $CP$ asymmetries for
the following $CP$-eigenstate $\bz$ decay modes:
$\bz\to \phi\kl$ and $\fzero\ks$ for $\xi_f = +1$;
$\bz\to\omega\ks$ 
and
$\ks\piz$
for $\xi_f = -1$.
The decays $\bz\to\phi\ks$ and $\phi\kl$ are combined
in this analysis by redefining $\cals$ as $-\xi_f\cals$ to take
the opposite $CP$ eigenvalues into account, 
and are collectively called ``$\bz\to\phi\kz$''.
The $CP$ asymmetries for the decay $\bz\to\omega\ks$
are measured for the first time.

At the KEKB energy-asymmetric 
$e^+e^-$ (3.5 on 8.0~GeV) collider~\cite{bib:KEKB},
the $\Upsilon(4S)$ is produced
with a Lorentz boost of $\beta\gamma=0.425$ nearly along
the electron beamline ($z$).
Since the $B^0$ and $\bzb$ mesons are approximately at 
rest in the $\Upsilon(4S)$ center-of-mass system (cms),
$\Delta t$ can be determined from the displacement in $z$ 
between the $f_{CP}$ and $f_{\rm tag}$ decay vertices:
$\Delta t \simeq (z_{CP} - z_{\rm tag})/(\beta\gamma c)
 \equiv \Delta z/(\beta\gamma c)$.

The Belle detector is a large-solid-angle magnetic
spectrometer that
consists of a silicon vertex detector (SVD),
a 50-layer central drift chamber (CDC), an array of
aerogel threshold Cherenkov counters (ACC),
a barrel-like arrangement of time-of-flight
scintillation counters (TOF), and an electromagnetic calorimeter
comprised of CsI(Tl) crystals (ECL) located inside
a superconducting solenoid coil that provides a 1.5~T
magnetic field.  An iron flux-return located outside of
the coil is instrumented to detect $K_L^0$ mesons and to identify
muons (KLM).  The detector
is described in detail elsewhere~\cite{Belle}.
Two inner detector configurations were used. A 2.0 cm radius beampipe
and a 3-layer silicon vertex detector (SVD-I) were used for DS-I,
while a 1.5 cm radius beampipe, a 4-layer
silicon detector (SVD-II) and a small-cell inner drift chamber were used for
DS-II~\cite{Ushiroda}.

%%%%%%%%%%%%%%%%%%%%%%%%%%%%%%%%%%%%%%%%%%%%%%%%%%%%%%%%%%%%%%%%%%%
\section{Event Selection, Flavor Tagging and Vertex Reconstruction}
%%%%%%%%%%%%%%%%%%%%%%%%%%%%%%%%%%%%%%%%%%%%%%%%%%%%%%%%%%%%%%%%%%%
%%%%%%%%%%%%%%%%%%%%%%%%%%%%%%%%%
\subsection{Overview}
%%%%%%%%%%%%%%%%%%%%%%%%%%%%%%%%%
We reconstruct the following $\bz$ decay modes to
measure $CP$ asymmetries:
$\bz\to\phi\ks$, 
$\phi\kl$, 
$\kp\km\ks$, 
$\fzero\ks$,
$\eta'\ks$,
$\omega\ks$, 
and
$\ks\piz$.
We exclude $K^+K^-$ pairs that are consistent with a $\phi \to K^+K^-$ decay 
from the $\bz \to K^+K^-\ks$ sample.
The intermediate meson states are reconstructed from the following decays:
$\piz\to\gamma\gamma$,
$\ks \to \pip\pim$ (also $\piz\piz$ for the $\phi\ks$ decay), 
$\eta\to\gamma\gamma$ or $\pip\pim\piz$, 
$\rhoz\to\pip\pim$,
$\omega\to\pip\pim\piz$,
$\eta'\to\rhoz\gamma$ or $\eta\pip\pim$,
$\fzero\to\pip\pim$,
and $\phi\to K^+K^-$.

Among the decay chains listed above,
$\bz\to\phi\ks~(\ks\to\pip\pim)$,
$\bz\to\kp\km\ks$,
$\bz\to\eta'\ks~(\eta'\to\rhoz\gamma)$, and
$\bz\to\eta'\ks~(\eta'\to\eta\pip\pim,~\eta\to\gamma\gamma)$
decays were used in the previous analysis~\cite{Abe:2003yt}.
The selection criteria for these decays remain the same.
%For the other $\bz$ decay modes that are now included,
For the newly included $\bz$ decay modes,
identification of photons,
neutral and charged kaons, and neutral and charged pions is based on
the procedure used previously.
However, the selection criteria
for each $\bz$ decay mode were optimized individually and
are thus different from one another.

%%%%%%%%%%%%%%%%%%%%%%%%%%%%%%%%%
\subsection{\boldmath $\bz\to\phi\ks$ and $\kp\km\ks$}
%%%%%%%%%%%%%%%%%%%%%%%%%%%%%%%%%
%We use well-reconstructed charged tracks with a sufficient
%number of associated hits in the CDC.
Charged tracks reconstructed with the CDC
for kaon and pion candidates except for those from $\ks\to\pip\pim$ decays
are required to originate from the interaction point (IP).
We distinguish charged kaons from pions based on
a kaon (pion) likelihood $\mathcal{L}_{K(\pi)}$
derived from the TOF, ACC and $dE/dx$ measurements in the CDC.

Pairs of oppositely charged tracks that have an invariant mass
within 0.030 GeV/$c^2$
of the nominal $\ks$ mass are used to reconstruct $\ks\to\pip\pim$ decays.
The distance of closest approach of the 
candidate charged tracks to the IP in the plane
perpendicular to $z$ axis is required to be larger than 0.02~cm
for high momentum ($>1.5$ GeV/$c$) $\ks$ candidates and 0.03~cm 
for those with momentum less than 1.5~GeV/$c$.
The $\pip\pim$ vertex is required to be displaced from
the IP by a minimum transverse distance of 0.22~cm for
high-momentum candidates and 0.08~cm for the remaining candidates.
The mismatch in the $z$ direction at the $\ks$ vertex point for the $\pip\pim$
tracks must be less than 2.4~cm for high-momentum candidates and 1.8~cm
for the remaining candidates.
The direction of the pion pair momentum must also agree with
the direction of the vertex point from the IP
to within 0.03 rad for high-momentum candidates, and 
to within 0.1 rad for the remaining candidates.
The resolution of the reconstructed $\ks$ mass is 0.003~GeV$/c^2$.
%The $\pip\pim$ vertex is required to be displaced from
%the IP by a minimum transverse distance of 0.22~cm for
%high momentum ($>1.5$ GeV/$c$) candidates and 0.08~cm for
%those with momentum less than 1.5~GeV/$c$.
%The direction of the pion pair momentum must also agree with
%the direction defined by the IP and the vertex displacement
%within 0.03 rad for high-momentum candidates, and within 0.1
%rad for the remaining candidates.

Photons are identified as isolated ECL clusters
that are not matched to any charged track.
To select $\ks\to\piz\piz$ decays,
we reconstruct $\piz$ candidates from pairs of photons
with $E_\gamma > 0.05$ GeV, where
$E_\gamma$ is the photon energy measured with the ECL.
Photon pairs with an invariant mass between
0.08 and 0.15 GeV$/c^2$ and a momentum above 0.1 GeV/$c$
are used as $\piz$ candidates.
Initially, the $\piz$ decay vertex is assumed to be the IP.
The asymmetric mass window is used to take into account the lower tail of
the mass distribution due to the distance between the IP and the
true $\piz$ vertex.
%The large mass range is used to achieve a high reconstruction efficiency.
Candidate $\ks\to\piz\piz$ decays are required to have
an invariant mass between 0.47 GeV/$c^2$ and 0.52 GeV/$c^2$,
where we perform a fit with constraints on
the $\ks$ vertex and the $\piz$ masses
to improve the $\piz\piz$ invariant mass resolution.
We also require that the distance between the IP and the
reconstructed $\ks$ decay vertex be larger than $-10$~cm,
where the positive direction is defined by the $\ks$ momentum.

Candidate $\phi \to \kp\km$ decays are required to
have an invariant mass that is within 0.01 GeV/$c^2$ 
of the nominal $\phi$ meson mass.
Since the $\phi$ meson selection is effective in reducing background events,
we impose only minimal kaon-identification requirements;
$\rkpi \equiv \mathcal{L}_K /(\mathcal{L}_K + \mathcal{L}_\pi) > 0.1$
is required, 
where the kaon likelihood ratio $\mathcal{R}_{K/\pi}$ has values 
between 0 (likely to be a pion) and 1 (likely to be a kaon).
We use a more stringent kaon-identification requirement,
$\rkpi > 0.6$,
to select non-resonant $\kp\km$ candidates 
for the decay $\bz \to \kp\km\ks$.
The $\kp\km$ candidates for $\bz \to \kp\km\ks$ are selected by rejecting
$\kp\km$ pairs with an invariant mass within 0.015 GeV/$c^2$ of the 
nominal $\phi$ meson mass, reducing the $\phi$ contribution to a negligible
level.
To remove $\chi_{c0}\to K^+K^-$, $J/\psi \to K^+K^-$ and $D^0\to K^+K^-$
decays, $K^+K^-$ pairs with an invariant mass within 0.015~GeV$/c^2$ of the 
nominal masses of $\chi_{c0}$ and $J/\psi$ or within 0.01~GeV$/c^2$ of the nominal 
$D^0$ mass are rejected.
$D^+ \to \ks K^+$ decays are also removed by rejecting $\ks K^+$
pairs with an invariant mass within 0.01~GeV$/c^2$ of the nominal $D^+$ mass.
%$D^0 \to K^+K^-$, $\chi_{c0} \to K^+K^-$, or $J/\psi \to K^+K^-$ decays.
%We also remove $D^+ \to \ks K^+$ candidates.

For reconstructed $B\to\fCP$ candidates, we identify $B$ meson decays using the
energy difference $\dE\equiv E_B^{\rm cms}-E_{\rm beam}^{\rm cms}$ and
the beam-energy constrained mass $\mb\equiv\sqrt{(E_{\rm beam}^{\rm cms})^2-
(p_B^{\rm cms})^2}$, where $E_{\rm beam}^{\rm cms}$ is
the beam energy in the cms, and
$E_B^{\rm cms}$ and $p_B^{\rm cms}$ are the cms energy and momentum of the 
reconstructed $B$ candidate, respectively.
The resolution of $\mb$ is about 0.003~GeV$/c^2$.
Because of the smallness of $p_B^{\rm cms}$,
the $\mb$ resolution is dominated by the beam-energy spread 
which is common to all decay modes.
The resolution of $\dE$ depends on the reconstructed decay mode.
The $\dE$ resolution is 
0.013~GeV for $\phi\ks~(\ks\to\pip\pim)$ and $\kp\km\ks$.
The $\dE$ distribution for $\phi\ks~(\ks\to\piz\piz)$ has a tail toward lower
$\dE$ 
due to $\gamma$ energy leaking in the ECL. 
%due to $\gamma$ energy leakage in ECL. The $\dE$ resolution
The $\dE$ resolution for $\phi\ks~(\ks\to\piz\piz)$ is 0.058~GeV 
for the main component
and the width of the tail component is about 0.14~GeV.
%and 74~MeV 
The $B$ meson signal region is defined as 
$|\dE|<0.06$ GeV for $\bz \to \phi \ks~(\ks\to\pip\pim)$,
$-0.15~{\rm GeV} < \dE < 0.1$ GeV for $\bz \to \phi \ks~(\ks\to\piz\piz)$,
$|\dE|<0.04$ GeV for $\bz \to K^+K^-\ks$,
and $5.27~{\rm GeV}/c^2 <\mb<5.29~{\rm GeV}/c^2$ for all decays.

The dominant background to the $\bz\to\phi\ks$ decay comes from
$e^+e^- \rightarrow 
u\overline{u},~d\overline{d},~s\overline{s}$, or $c\overline{c}$
continuum events. Since these tend to be jet-like, while
the signal events tend to be spherical,
we use a set of variables that characterize the event topology
to distinguish between the two.
We combine 
$\sperp$, $\theta_T$ and modified Fox-Wolfram moments~\cite{Abe:2001nq}
into a Fisher discriminant $\calf$, where
$\sperp$ is
the scalar sum of the transverse momenta of 
%all particles outside a $45^\circ$ cone around the candidate $\phi$
particles other than the reconstructed $B$ candidate
outside a $45^\circ$ cone around the 
candidate $\phi$ meson direction 
(the thrust axis of the $B$ candidate for $\kp\km\ks$ decays)
divided by the scalar sum of their total momenta, and $\theta_T$ is
the angle between the thrust axis of the $B$ candidate
and that of the other particles in the cms.
We also use the angle of the reconstructed $B$
candidate with respect to the beam direction in the cms
($\theta_B$), 
and the helicity angle $\theta_H$ defined as the
angle between the $B$ meson momentum and the daughter
$\kp$ momentum in the $\phi$ meson rest frame.
We combine
$\calf$, $\cos\theta_B$ and $\cos\theta_H$ 
into a signal [background]
likelihood variable, which is defined as 
${\cal L}_{\rm sig[bkg]} \equiv
{\cal L}_{\rm sig[bkg]}(\calf)\times
{\cal L}_{\rm sig[bkg]}(\cos\theta_B)\times
{\cal L}_{\rm sig[bkg]}(\cos\theta_H)$.
For $\kp\km\ks$ decays, $\calf$ and $\cos\theta_B$ are 
combined to make the likelihood variables.
We impose requirements on the likelihood ratio
$\rsigbkg \equiv \lsig/(\lsig+\lbkg)$ to
maximize the figure-of-merit (FoM) defined as
$\nsigmc/\sqrt{\nsigmc+\nbkg}$, where $\nsigmc$ ($\nbkg$) 
represents the expected
number of signal (background) events in the signal region.
We estimate $\nsigmc$ using Monte Carlo (MC) events, while
$\nbkg$ is determined from events outside the signal region.
The requirement for $\rsigbkg$
depends both on the decay mode
and on the flavor-tagging quality, $r$, which is
described in Sec.~\ref{sec:flavor tagging}.
The threshold values range from 0.1 (used for $r>0.875$)
to 0.4 (used for $r<0.25$) for the decay
$\bz\to\phi\ks~(\ks\to\pip\pim)$,
and from 0.25 to 0.65 for the decay $\bz\to\kp\km\ks$.
For the $\bz\to\kp\km\ks$ candidates, we also require $|\cos\theta_T|<0.9$ 
prior to the $\rsigbkg$ requirement.
We impose a more stringent requirement, $\rsigbkg > 0.75$,
for all $r$ values
in the decay $\bz\to\phi\ks~(\ks\to\piz\piz)$.
The $\rsigbkg$ requirement reduces the continuum background
by 73\% for $\bz\to\phi\ks~(\ks\to\pip\pim)$, 
92\% for $\bz\to\kp\km\ks$ and 
94\% for $\bz\to\phi\ks~(\ks\to\piz\piz)$,
retaining 91\% of the signal 
     for $\bz\to\phi\ks~(\ks\to\pip\pim)$,
72\% for $\bz\to\kp\km\ks$ and 
71\% for $\bz\to\phi\ks~(\ks\to\piz\piz)$.

We use events outside the signal region 
as well as a large MC sample to study the background components.
The dominant background is from continuum.
The contributions from $B\overline{B}$ events are small.
We estimate the contamination of $\bz\to\kp\km\ks$ and 
$\bz\to\fzero\ks~(\fzero\to\kp\km)$ decays in the $\bz\to\phi\ks$ sample
from the $B$ yields in $\kp\km$ mass sideband data.
The contamination of $\bz\to\kp\km\ks$ events in the $\bz\to\phi\ks$ sample 
is $7.1\pm1.6$\% ($6.2\pm2.0$\%) for DS-I (DS-II).
Backgrounds from the decay $\bz\to\fzero\ks~(\fzero\to\kp\km)$, 
which has a
$CP$ eigenvalue opposite to $\phi\ks$, are found to be 
$0.4^{+1.9}_{-0.4}$\% ($0.0^{+2.0}_{-0.0}$\%) for DS-I (DS-II).
The influence of these backgrounds
is treated as a source of systematic uncertainty.

\begin{figure}
\includegraphics[width=0.32\textwidth]{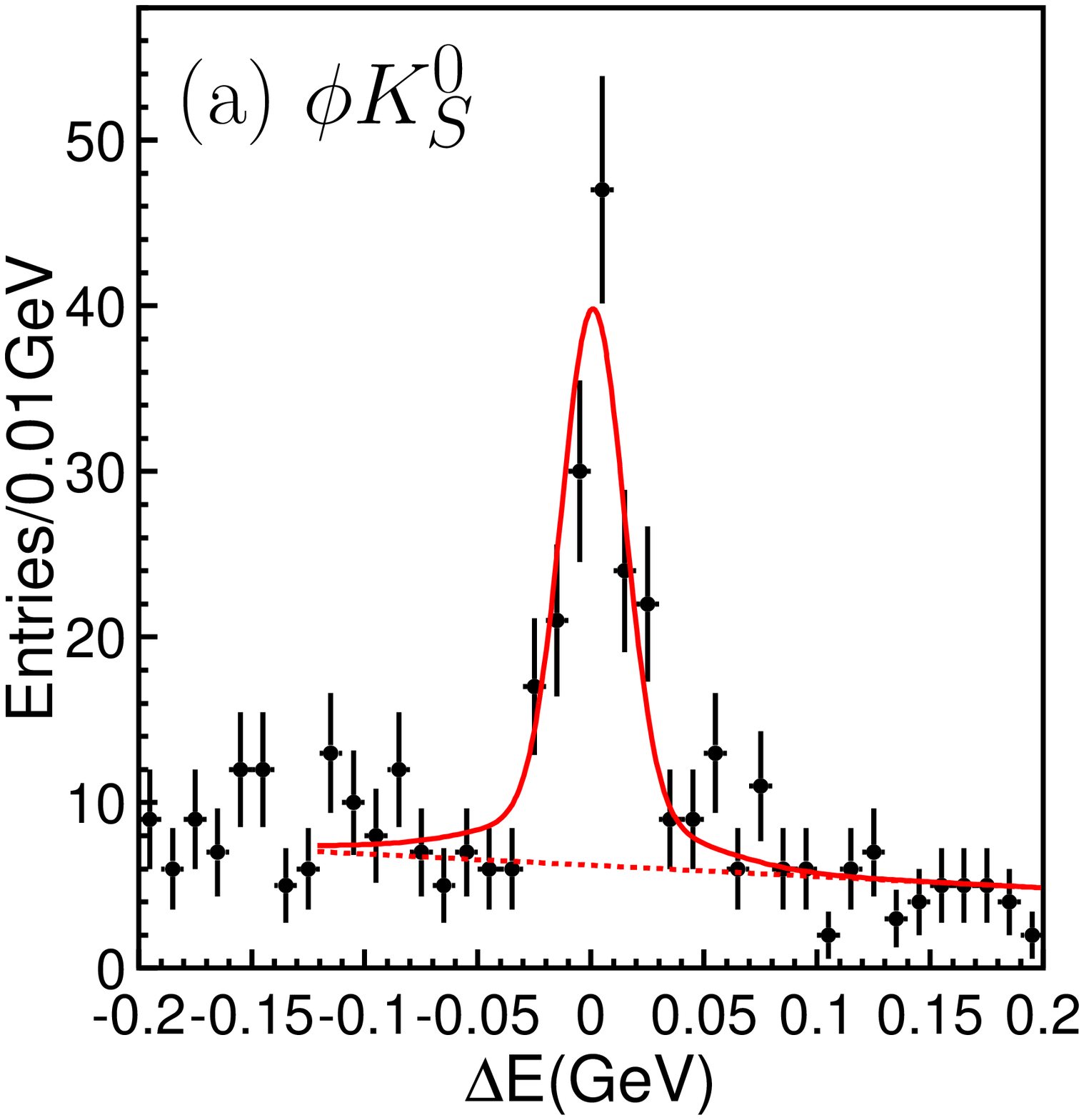}
\includegraphics[width=0.32\textwidth]{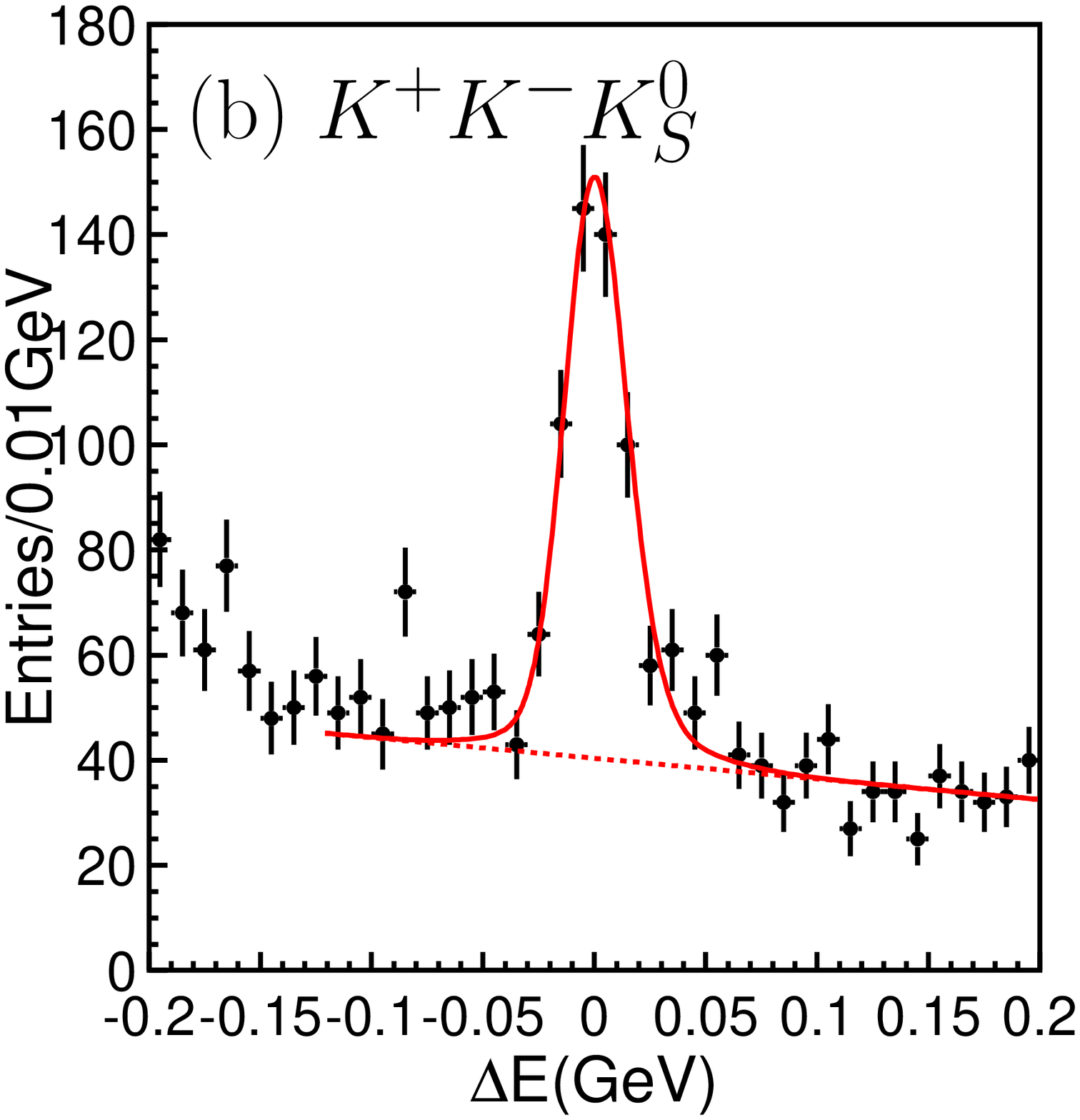}
\includegraphics[width=0.32\textwidth]{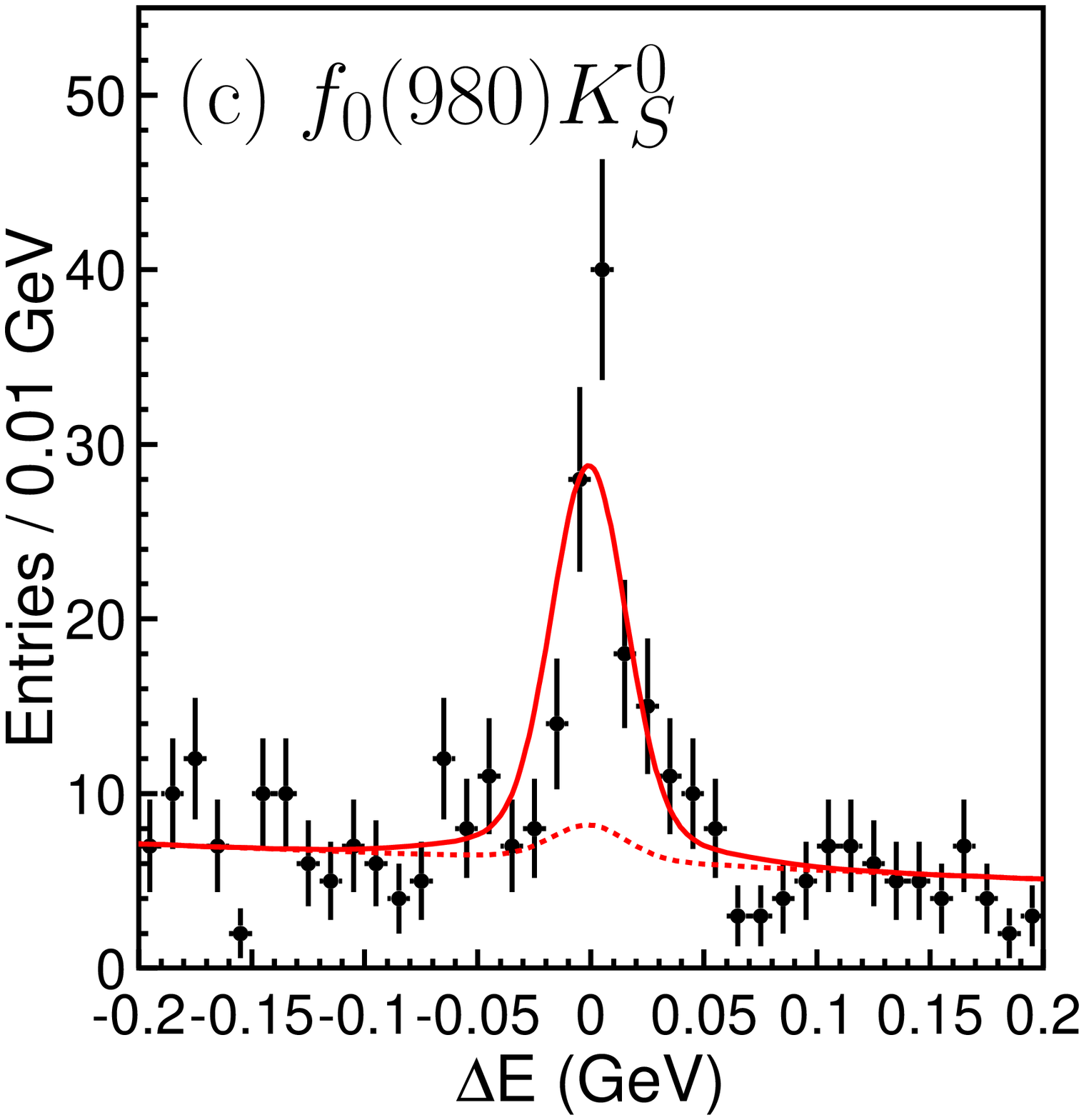}
\includegraphics[width=0.32\textwidth]{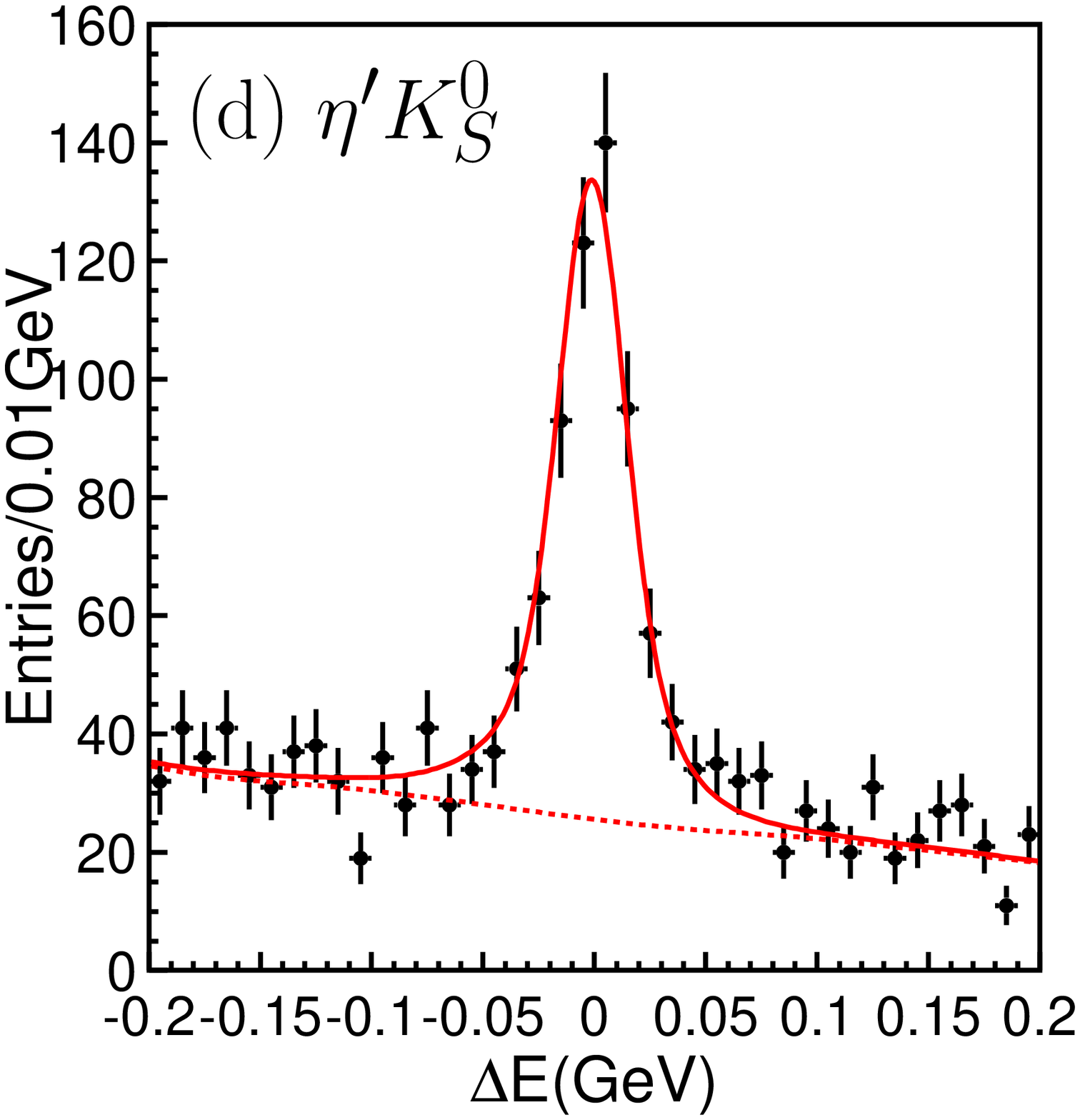}
\includegraphics[width=0.32\textwidth]{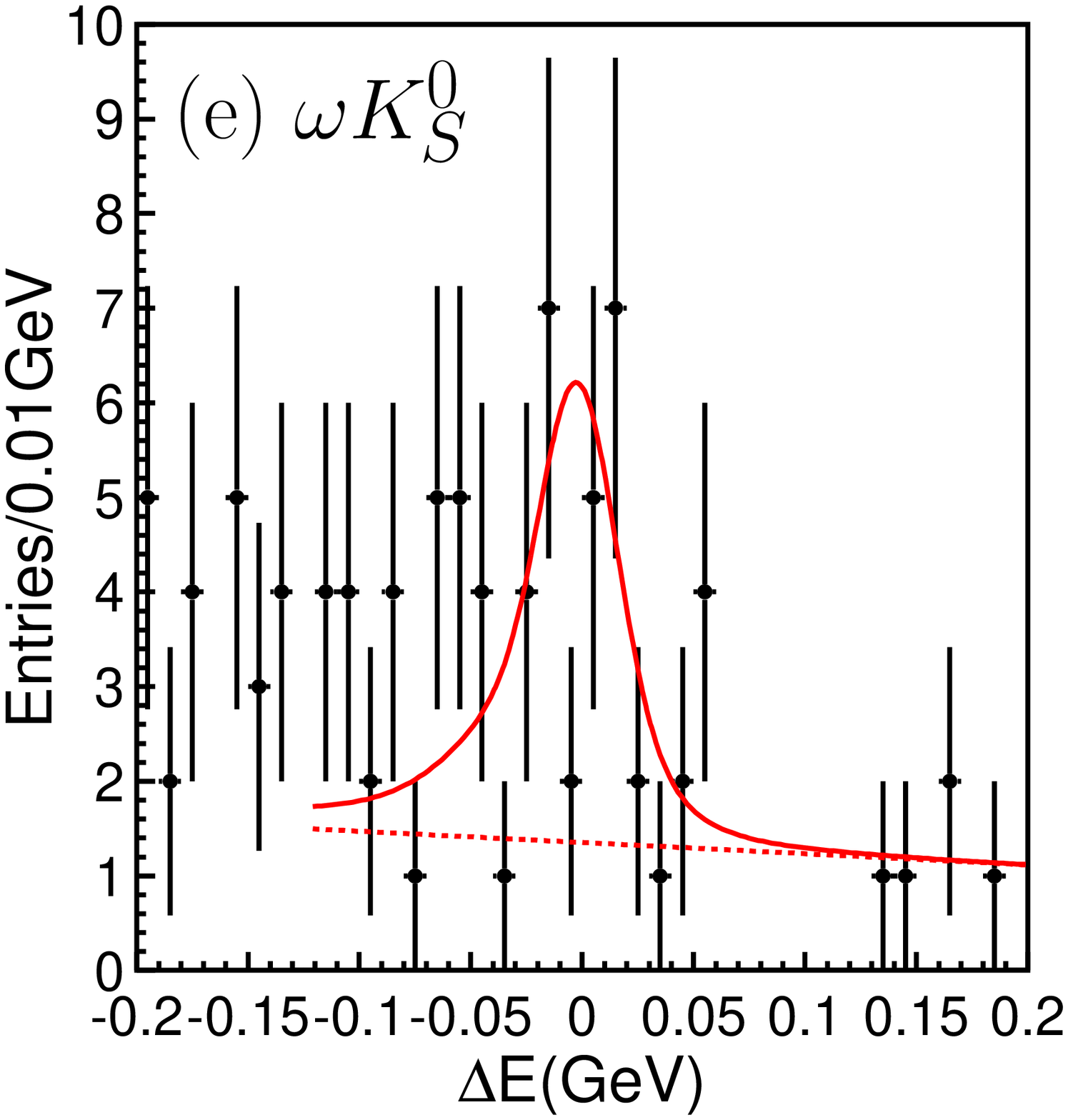}
\includegraphics[width=0.32\textwidth]{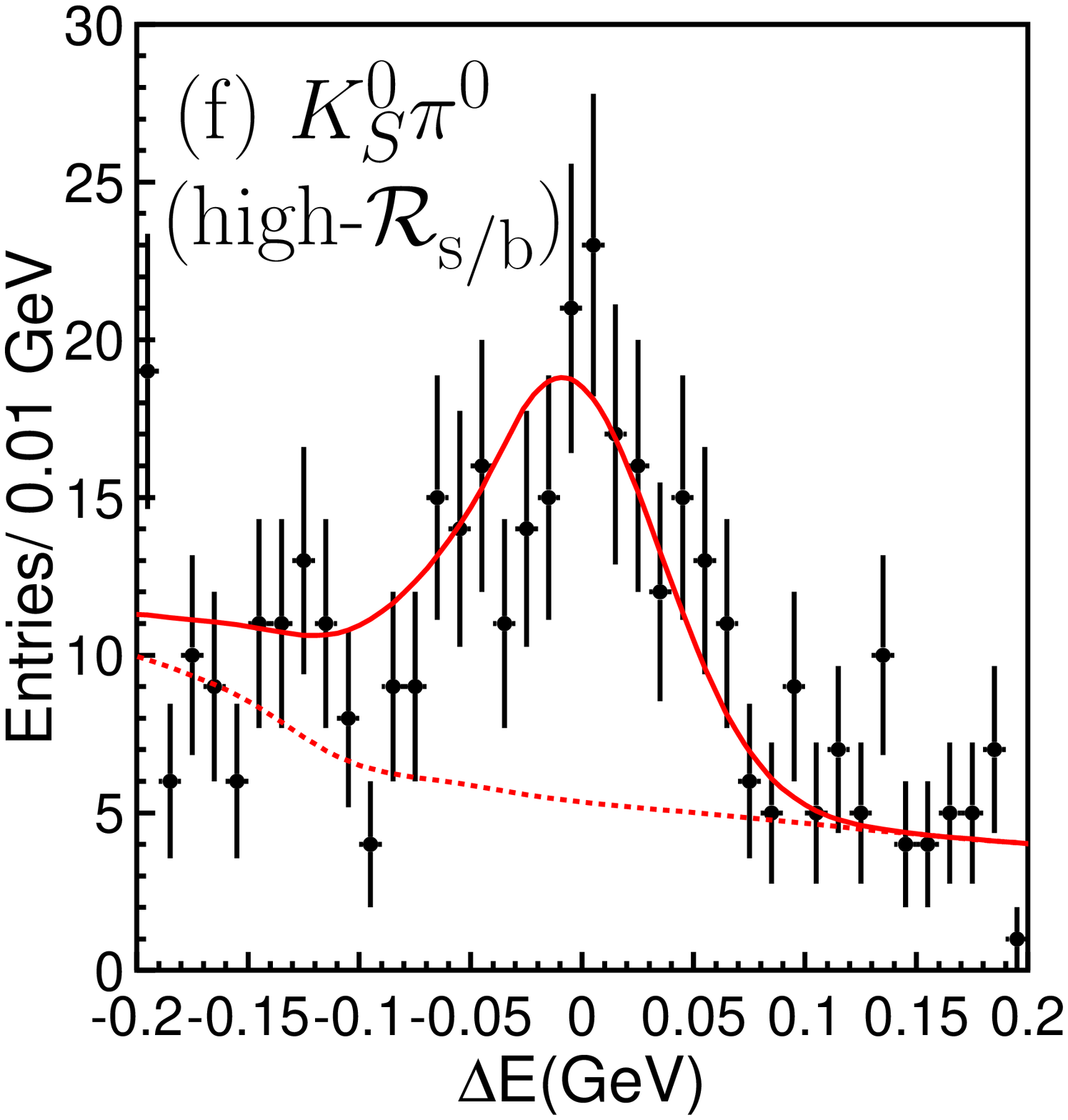}
\includegraphics[width=0.32\textwidth]{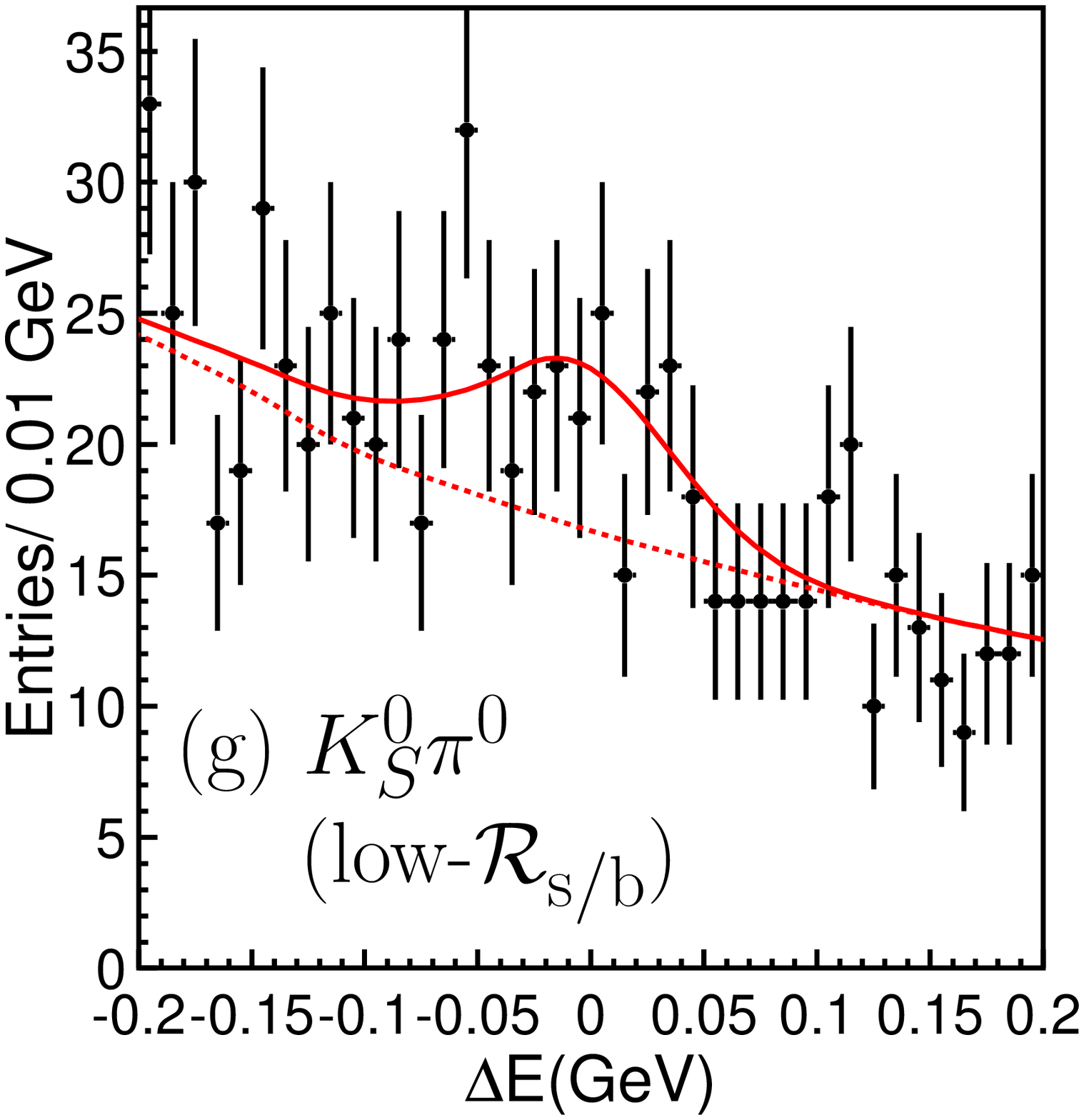}
\caption{$\dE$ distributions for
(a) $\bz\to\phi\ks$, (b) $\bz\to\kp\km\ks$, (c) $\bz\to\fzero\ks$,
(d) $\bz\to\eta'\ks$, (e) $\bz\to\omega\ks$, (f) $\bz\to\ks\piz$ 
(high-$\rsigbkg$), 
and (g) $\bz\to\ks\piz$ (low-$\rsigbkg$), 
within the $\mb$ signal region.
Solid curves show the fit to signal plus background distributions,
and dashed curves show the background contributions.
}\label{fig:de}
%
%\rput[l](-6.5, 17.0)  {(a)~$\phi\ks$}
%\rput[l](-1.4, 17.0)  {(b)~$\phi\kl$}
%\rput[l]( 3.65,17.0)  {(c)~$\kp\km\ks$}
%\rput[l](-6.5, 12.0)  {(d)~$\fzero\ks$}
%\rput[l](-1.4, 12.0)  {(e)~$\eta'\ks$}
%\rput[l]( 3.65,12.0)  {(f)~$\omega\ks$}
%\rput[l](-4.0,  7.0)  {(g)~$\ks\piz$}
%\rput[l](-3.5,  6.5)  {(high-$\rsigbkg$)}
%\rput[l]( 1.0,  4.5)  {(h)~$\ks\piz$}
%\rput[l]( 1.5,  4.0)  {(low-$\rsigbkg$)}
%
\end{figure}

Figures~\ref{fig:mb}(a) and (c) show
the $\mb$ distributions for the reconstructed $\bz\to\phi\ks$
and $\kp\km\ks$ candidates
within the $\dE$ signal regions
after flavor tagging and vertex reconstruction.
The $\dE$ distributions for the $\bz\to\phi\ks$ and $\kp\km\ks$ candidates 
within the $\mb$ signal region are shown in Fig.~\ref{fig:de}(a) and (b), 
respectively.
The signal yield is determined
from an unbinned two-dimensional maximum-likelihood fit
to the $\dE$-$\mb$ distribution in the fit region
defined as
$\mb > 5.2~{\rm GeV/}c^2$
for all modes, and
$-0.12~{\rm GeV} < \dE < 0.25~{\rm GeV}$ for the
$\bz\to\phi\ks~(\ks\to\pip\pim)$ or $\kp\km\ks$ decay and
$-0.15~{\rm GeV} < \dE < 0.25~{\rm GeV}$ for the
$\bz\to\phi\ks~(\ks\to\piz\piz)$ decay.
The $\phi\ks~(\ks\to\pip\pim)$ signal distribution 
is modeled with a Gaussian function (a sum of two Gaussian functions)
for $\mb$ ($\dE$).
The $\phi\ks~(\ks\to\piz\piz)$ signal distribution 
is modeled with a smoothed histogram obtained from MC events.
For the continuum background,
we use the ARGUS parameterization~\cite{bib:ARGUS} 
for $\mb$
and a linear function for $\dE$.
The fits yield $\Nsigphiks$ $\bz\to\phi\ks$ events and
$\Nsigkpkmks$ $\bz\to\kp\km\ks$ events in the signal region,
where the errors are statistical only.

%%%%%%%%%%%%%%%%%%%%%%%%%%%%%%%%%
\subsection{\boldmath $\bz\to\phi\kl$}
\label{sec:bztophikl}
%%%%%%%%%%%%%%%%%%%%%%%%%%%%%%%%%
Candidate $\phi \to \kp\km$ decays
are selected with the criteria described above. 
We select $\kl$ candidates based on KLM and ECL
information. There are two classes of $\kl$ candidates, 
which we refer to as KLM and ECL candidates.
The requirements for the KLM candidates are the same
as those used in the $\bz\to\jpsi\kl$ selection
for the $\sinbb$ measurement~\cite{bib:CP1_Belle}.
ECL candidates are selected from ECL clusters using
a $\kl$ likelihood ratio~\cite{bib:CP1_Belle}, which is
calculated from
the following information: the distance between the
ECL cluster and the closest extrapolated charged track position;
the ECL cluster energy; $E_9/E_{25}$, the ratio of energies summed in
$3\times 3$ and $5\times 5$ arrays of CsI(Tl) crystals surrounding
the crystal at the center of the shower; the ECL shower
width and the invariant mass of the shower. 
The likelihood ratio is required to be greater than 0.8.
For both KLM and ECL candidates, we also require
that the cosine of the angle between the $\kl$ direction
and the direction of the missing momentum of the event
in the laboratory frame be greater than 0.6.

Since the energy of the $\kl$ is not measured,
$\mb$ and $\dE$ cannot be calculated
in the same way that is used for the other final states.
Using the four-momentum of a reconstructed
$\phi$ candidate and the $\kl$ flight direction,
we calculate the momentum of the $\kl$ candidate
requiring $\dE=0$. We then calculate
$\pbstar$, the momentum of the $B$ candidate in the
cms, and define the
$B$ meson signal region as 
$0.2~{\rm GeV/}c < \pbstar < 0.5~{\rm GeV/}c$.
We impose the requirement $\rsigbkg > 0.98$ to reduce the continuum background
by 99.4\%. The signal efficiency of the $\rsigbkg$ requirement is 28.1\%.
Here $\rsigbkg$ is based on the discriminating variables
used for the $\bz\to\phi\ks$ decay and the number of
tracks originating from the IP with a momentum above 0.1 GeV/$c$.
The $\rsigbkg$ requirement is chosen to optimize the FoM, which is
calculated taking the background from
both continuum and generic $B$ decays into account.
The $\kl$ detection efficiency difference between data and MC 
is studied using the decay $\bz\to\jpsi\kl$, and corrections
are applied to the $\bz\to\phi\kl$ MC events to calculate
the FoM.
If there is more than one candidate $\bz\to\phi\kl$ decay
in the signal region,
we take the one with the highest $\rsigbkg$ value.
ECL candidates are not used if there is a candidate
$\bz\to\phi\kl$ decay with a KLM candidate.
We find that about 90\% of signal events are reconstructed
with KLM candidates.

We study the background components
using a large MC sample
as well as data taken with cms energy 60~MeV
below the nominal $\Upsilon(4S)$ mass (off-resonance data).
The dominant background is from continuum.
A MC study with the efficiency correction obtained from
$\bz\to\jpsi\kl$ data yields
$9\pm 5$ background events
from $B$ decays, which include
$\bz\to\phi\kstarz$, $\phi\ks$ and $\bp\to\phi\kstarp$
decays.
The influence of these backgrounds, including
their $CP$ asymmetries,  is treated as a source
of systematic uncertainty.

The $\pbstar$ distribution after flavor tagging 
and vertex reconstruction
is shown in Fig.~\ref{fig:mb}(b).
The signal yield is determined from
an extended unbinned maximum-likelihood fit in the range
%$0~{\rm GeV/}c < \pbstar < 1~{\rm GeV/}c$.
$\pbstar < 1~{\rm GeV/}c$.
The $\bz\to\phi\kl$ signal shape is obtained from MC events.
Background from $B\bbar$ pairs is also modeled with
MC. We fix the ratio between the signal yield
and the $B\bbar$ background based on
known branching fractions and reconstruction
efficiencies; the uncertainty in the ratio is treated
as a source of systematic error.
The continuum background distribution is represented
by a smoothed histogram obtained from MC events;
we confirm that the function describes
the off-resonance data well.
The fit yields $\Nsigphikl\pm 10$ $\bz\to\phi\kl$ events,
where the first error is statistical and the second
error is systematic.
The sources of the systematic error include
uncertainties in the efficiency corrections,
in $B\bbar$ background branching fractions and
in the background parameterizations.
The result is in good agreement with
the expected $\bz\to\phi\kl$ signal yield 
($36\pm 9$ events) obtained from MC after applying
the efficiency correction from the
$\bz\to\jpsi\kl$ data.

%%%%%%%%%%%%%%%%%%%%%%%%%%%%%%%%%
\subsection{\boldmath $\bz\to\fzero\ks$}
%%%%%%%%%%%%%%%%%%%%%%%%%%%%%%%%%
%Candidate $\ks\to\pip\pim$ decays are selected
%with the criteria that are slightly different from
%those used for the $\bz\to\phi\ks$ decay
%to obtain the best performance for the $\bz\to\fzero\ks$ decay. 
Candidate $\ks\to\pip\pim$ decays are selected
with criteria that are slightly different from
those used for the $\bz\to\phi\ks$ decay
so as to obtain the best performance for the $\bz\to\fzero\ks$ decay.
%The distance from the candidate charged tracks from the IP in the plane
%perpendicular to $z$ axis is required to be larger than 0.005~cm.
%The displacement of the $\pip\pim$ vertex from the IP is required to be 
%larger than 0.1~cm.
%A mismatch in the $z$ direction at the $\ks$ vertex point for the $\pip\pim$
%tracks must be less than 3~cm.
%The difference between the direction of the pion pair momentum 
%and the direction defined by the IP and the vertex displacement
%is required to be within 0.02~rad.
Pairs of oppositely charged pions that have invariant masses
between 0.890 and 1.088 GeV/$c^2$ are used to reconstruct
$\fzero\to\pip\pim$ decays.
Tracks that are identified as kaons ($\rkpi > 0.7$) or
electrons are not used.
We require that both $\ks\pip$ and $\ks\pim$
combinations have invariant masses more than
0.1 GeV/$c^2$ above the nominal charged
$D$ meson mass; this
removes background from $D^\pm\to\ks\pi^\pm$ and 
$K^{*\pm}\to\ks\pi^\pm$ decays.

The $B$ meson signal region is defined as
$|\dE|<0.06$ GeV and $5.27~{\rm GeV}/c^2 <\mb<5.29~{\rm GeV}/c^2$.
The $\dE$ resolution is 0.019~GeV.
The dominant background is from continuum.
For the continuum suppression, we require
$\rsigbkg > 0.6$ for events with the best-quality flavor tagging
($r > 0.875$), and $\rsigbkg > 0.8$ for other events.
The continuum background is reduced by 95\%, retaining 60\% of signal events.
Here the signal likelihood ratio $\rsigbkg$ is obtained from
$\cos\theta_B$ and
$\calf$, which consists of the modified Fox-Wolfram moments
and $\cos\theta_T$.

\begin{figure}
\includegraphics[width=0.75\textwidth]{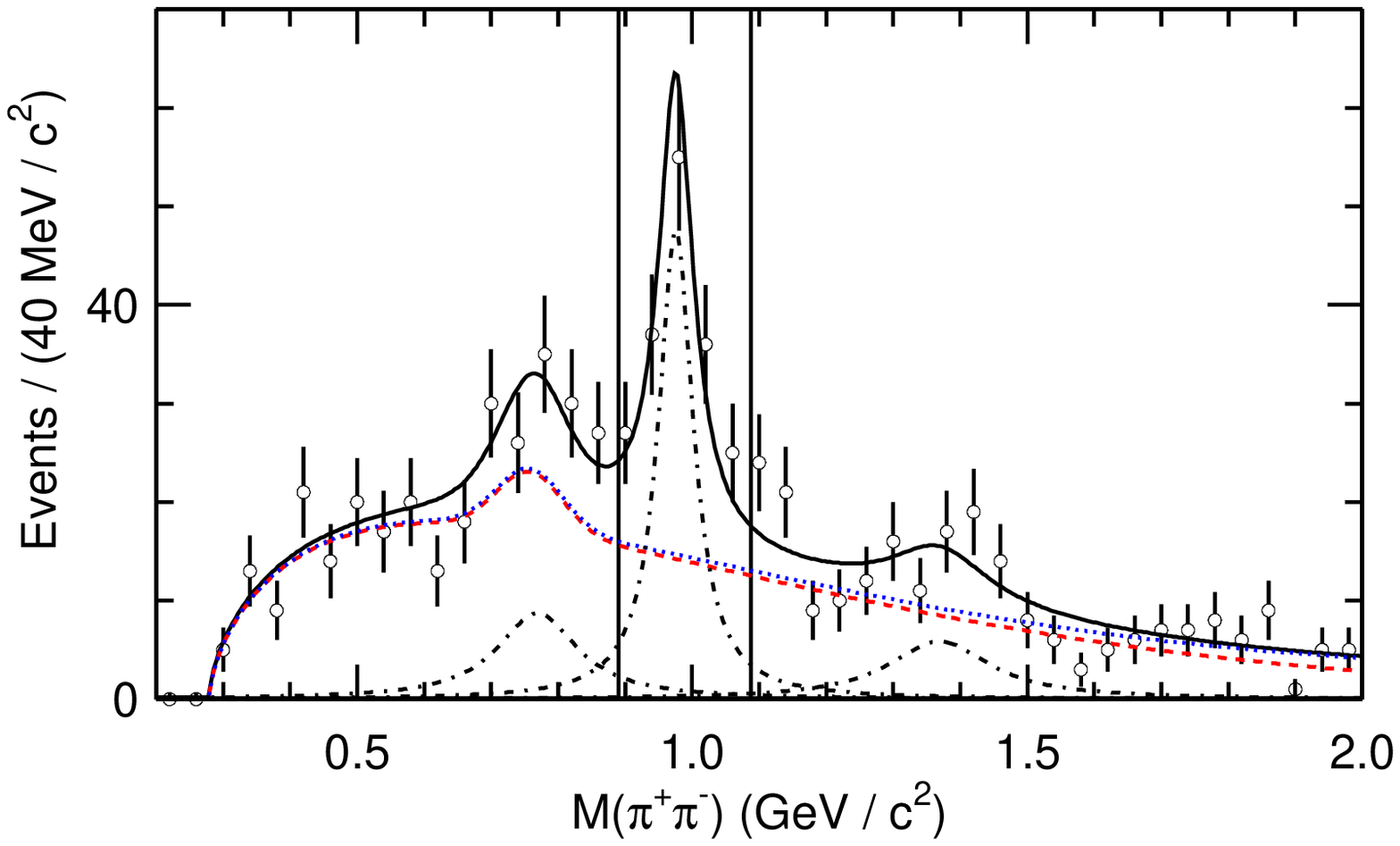}
\caption{
Invariant mass distribution of $\pip\pim$ pairs for $\bz\to\fzero\ks$
candidates in the $\dE$-$\mb$ signal region.
The solid curve shows the fit result.
The dashed line shows continuum background.
The dotted line shows the sum of the continuum and 
the non-resonant $\bz\to\pip\pim\ks$.
The dot-dashed lines show the $\bz\to\rhoz\ks$ (left),
the $\bz\to\fzero\ks$ (middle), and the $\bz\to\fx\ks$ (right).
}\label{fig:mpp}
\end{figure}

Figure~\ref{fig:mb}(d) shows
the $\mb$ distribution for the reconstructed $\bz\to\fzero\ks$ candidates
within the $\dE$ signal region
after flavor tagging and vertex reconstruction.
The $\dE$ distribution for the $\bz\to\fzero\ks$ candidates 
within the $\mb$ signal region is shown in Fig.~\ref{fig:de}(c).
For the signal yield extraction,
we first perform
an unbinned two-dimensional maximum-likelihood fit
to the $\dE$-$\mb$ distribution in the fit region defined as
$\mb > 5.2~{\rm GeV/}c^2$
and
$-0.3~{\rm GeV} < \dE < 0.4~{\rm GeV}$.
The signal is modeled with a Gaussian function
(a sum of two Gaussian functions) for $\mb$ ($\dE$).
For the continuum background, 
we use the ARGUS parameterization for $\mb$
and a linear function for $\dE$.
The fit yields the number of 
$\bz\to\pip\pim\ks$ events that have
$\pip\pim$ invariant masses within
the $\fzero$ resonance region, which
may include contributions from
$\bz\to\rhoz\ks$ as well as non-resonant three-body
$\bz\to\pip\pim\ks$ decays.
To separate these peaking backgrounds from
the $\bz\to\fzero\ks$ decay,
we perform another fit to the $\pip\pim$ invariant
mass distribution for the events inside the
$\dE$-$\mb$ signal region. We use Breit-Wigner
functions for the $\bz\to\fzero\ks$ signal as well
as for $\bz\to\rho\ks$ and a possible resonance
above the $\fzero$ mass region, which is referred to as
$\fx$. We use the masses and widths of $\fzero$ and $\fx$ obtained
from data~\cite{Garmash:2003er}.
Three-body $\bz\to\pip\pim\ks$ decays are modeled with
a fourth-order polynomial function. 
%The other background is modeled with a threshold function.
The continuum background is modeled with a sum of 
a threshold function and a Breit-Wigner function for $\rho$ resonance.
The $\pip\pim$ invariant mass distribution with the fit result
is shown in Fig.~\ref{fig:mpp}.
The fit yields $\Nsigfzeroks$ $\bz\to\fzero\ks$ events.
The peaking background contribution in the 
$\dE$-$\mb$ signal region is estimated to be $9\pm 3$ events.

%%%%%%%%%%%%%%%%%%%%%%%%%%%%%%%%%
\subsection{\boldmath $\bz\to\eta'\ks$}
%%%%%%%%%%%%%%%%%%%%%%%%%%%%%%%%%
Candidate $\ks \to \pip\pim$ decays
are selected with the same criteria as those used for
the $\bz\to\phi\ks$ decay.
Charged pions from the $\eta$, $\rhoz$ or $\eta'$ decay
are selected from tracks originating from the IP.
We reject kaon candidates by requiring $\rkpi < 0.9$. 
Candidate photons from
$\piz\to\gamma\gamma$ decays are
required to have $E_\gamma > 0.05 $~GeV.
The reconstructed $\piz$ candidate is required to satisfy 
$0.118~{\rm GeV}/c^2 < \mgg < 0.15~{\rm GeV}/c^2$
and
$\ppizcms > 0.1~{\rm GeV}/c$, where
$\mgg$ and $\ppizcms$ are the invariant mass and
the momentum in the cms, respectively. 
Candidate photons from
$\eta\to\gamma\gamma~(\eta'\to\rhoz\gamma)$ decays are
required to have $E_\gamma > 0.05~(0.1)$~GeV.
The invariant mass of the photon pair
is required to be between 0.5 and
0.57~GeV/$c^2$ for the $\eta\to\gamma\gamma$ decay.
The $\pip\pim\piz$ invariant mass is required
to be between 0.535 and 0.558~GeV/$c^2$ for the
$\eta\to\pip\pim\piz$ decay.
A kinematic fit with an $\eta$ mass constraint is
performed using the fitted vertex of the $\pi^+\pi^-$ tracks from
the $\eta^\prime$ as the decay point. 
For $\eta^\prime\to\rhoz\gamma$ decays, candidate $\rhoz$ mesons
are reconstructed from pairs of vertex-constrained $\pi^+\pi^-$
tracks with an invariant mass between 0.55 and 0.92~GeV/$c^2$. 
The $\eta^\prime\to\eta\pip\pim$ candidates are required 
to have a reconstructed mass
between 0.94 and 0.97~GeV/$c^2$ (0.95 and 0.966~GeV/$c^2$)
for the $\eta\to\gamma\gamma$ ($\eta\to\pip\pim\piz$) decay.
Candidate $\eta^\prime\to\rhoz\gamma$ decays are required to
have a reconstructed mass from 0.935 to 0.975~GeV/$c^2$.

The $B$ meson signal region is defined as 
$|\dE|<0.06$ GeV for $\bz \to \eta'\ks~(\eta'\to \rhoz\gamma) $,
$-0.1$ GeV $< \dE <0.08$ GeV for 
$\bz \to \eta'\ks~(\eta'\to\eta\pip\pim,~\eta\to\gamma\gamma) $,
$-0.08$ GeV $< \dE <0.06$ GeV for 
$\bz \to \eta'\ks~(\eta'\to\eta\pip\pim,~\eta\to\pip\pim\piz) $,
and $5.27~{\rm GeV}/c^2 <\mb<5.29~{\rm GeV}/c^2$ for all decays.
The $\dE$ resolution is 0.017~GeV for $\eta'\to\rhoz\gamma$, 
0.027~GeV for $\eta'\to\eta\pip\pim~(\eta\to\gamma\gamma)$
and 0.018~GeV for $\eta'\to\eta\pip\pim~(\eta\to\pip\pim\piz)$.
The continuum suppression is based on the
likelihood ratio $\rsigbkg$ obtained from
the same discriminating variables 
used for the $\bz\to\phi\ks$ decay, except that
we only use $\cos\theta_H$ for the decay 
$\eta'\to\rho\gamma~(\rho\to\pip\pim)$, where $\theta_H$ is defined as
the angle between the $\eta'$ meson momentum and the daughter
$\pip$ momentum in the $\rho$ meson rest frame.
The minimum $\rsigbkg$ requirement
depends both on the decay mode and on the
flavor-tagging quality, and
ranges from 0 (i.e., no requirement) to $0.4$.
For the $\eta'\to\rhoz\gamma$ mode, we also require $|\cos\theta_T|<0.9$ prior
to the $\rsigbkg$ requirement.
With these requirements, the continuum background is reduced by  
87\% for $\eta'\to\rhoz\gamma$,
58\% for $\eta'\to\eta\pip\pim~(\eta\to\gamma\gamma)$ and
31\% for $\eta'\to\eta\pip\pim~(\eta\to\pip\pim\piz)$,
retaining 78\% of the signal
     for $\eta'\to\rhoz\gamma$,
94\% for $\eta'\to\eta\pip\pim~(\eta\to\gamma\gamma)$ and
97\% for $\eta'\to\eta\pip\pim~(\eta\to\pip\pim\piz)$

We use events outside the signal region 
as well as a large MC sample to study the background components.
The dominant background is from continuum.
In addition, according to MC simulation, there is  a small ($\sim 3\%$) 
%contamination from $B\overline{B}$ background events
combinatorial background from $B\overline{B}$ events
in $\bz\to\eta'\ks~(\eta'\to\rhoz\gamma)$. 
The contributions from $B\overline{B}$ events are smaller for other modes.
The influence of these backgrounds
is treated as a source of systematic uncertainty.

Figure~\ref{fig:mb}(e) shows
the $\mb$ distribution for the reconstructed $\bz\to\eta'\ks$ candidates
within the $\dE$ signal region
after flavor tagging and vertex reconstruction,
where all subdecay modes are combined.
The $\dE$ distribution for the $\bz\to\eta'\ks$ candidates 
within the $\mb$ signal region is shown in Fig.~\ref{fig:de}(d).
The signal yields are determined
from unbinned two-dimensional maximum-likelihood fits
to the $\dE$-$\mb$ distributions in the fit region defined as
$\mb > 5.2~{\rm GeV/}c^2$
and
$-0.25~{\rm GeV} < \dE < 0.25~{\rm GeV}$.
We perform the fit for each final state separately.
The $\eta'\ks$ signal distribution 
is modeled with a sum of two (three) Gaussian functions
for $\mb$ ($\dE$).
For the continuum background, 
we use the ARGUS parameterization for $\mb$
and a linear function for $\dE$.
For the $\eta'\to\rho\gamma$ mode, we include
the $B\bbar$ background shape obtained from MC
in the fits.
The fits yield a total of $\Nsigetapks$ $\bz\to\eta'\ks$ events
in the signal region,
where the error is statistical only.

%%%%%%%%%%%%%%%%%%%%%%%%%%%%%%%%%
\subsection{\boldmath $\bz\to\omega\ks$}
%%%%%%%%%%%%%%%%%%%%%%%%%%%%%%%%%
Candidate $\ks \to \pip\pim$ decays
are selected with criteria that
are identical to those used for the $\bz\to\phi\ks$ decay.
Pions for the $\omega\to\pip\pim\piz$ decay
are selected with the same criteria used
for the $\eta\to\pip\pim\piz$ decay, except that
we require $\ppizcms > 0.35~{\rm GeV}/c$.
The $\pip\pim\piz$ invariant mass is required to be
within 0.03 GeV/$c^2$ of the nominal $\omega$ mass.
The $B$ meson signal region is defined as 
%$|\dE|<0.06$ GeV and $5.27~{\rm GeV}/c^2 <\mb<5.29~{\rm GeV}/c^2$.
$-0.10~{\rm GeV}<\dE<0.08~{\rm GeV}$ GeV 
and $5.27~{\rm GeV}/c^2 <\mb<5.29~{\rm GeV}/c^2$.
The $\dE$ resolution is 0.028~GeV.
The dominant background is from continuum.
The continuum suppression is based on the
likelihood ratio $\rsigbkg$ obtained from
the same discriminating variables 
used for the $\bz\to\phi\ks$ decay;
the helicity angle $\theta_H$ is defined as the
angle between the $\bz$ meson momentum and the cross product
of the $\pip$ and $\pim$ momenta
in the $\omega$ meson rest frame.
We also require $|\cos\theta_T|<0.9$ prior to the $\rsigbkg$ requirement.
The minimum $\rsigbkg$ requirement
%depends both on the decay mode and on the <--- ???
depends on the flavor-tagging quality, and
ranges from 0.3 (used for $r>0.875$)
to 0.9 (used for $r<0.25$).
The $\rsigbkg$ and $|\cos\theta_T|$ requirements reject 98\% of
the continuum background while retaining 56\% of the signal.
The contribution from $B\overline{B}$ events is negligibly small.

Figure~\ref{fig:mb}(f) shows
the $\mb$ distribution for the reconstructed $\bz\to\omega\ks$ candidates
within the $\dE$ signal region
after flavor tagging and vertex reconstruction.
The $\dE$ distribution for the $\bz\to\omega\ks$ candidates 
within the $\mb$ signal region is shown in Fig.~\ref{fig:de}(e).
The signal yield is determined
from an unbinned two-dimensional maximum-likelihood fit
to the $\dE$-$\mb$ distribution in the fit region defined as
$\mb > 5.2~{\rm GeV/}c^2$
and
$-0.12~{\rm GeV} < \dE < 0.25~{\rm GeV}$.
The signal distribution is modeled with a sum of two (three) Gaussian
functions for $\mb$ ($\dE$).
For the continuum background, 
we use the ARGUS parameterization for $\mb$
and a linear function for $\dE$.
The fit yields $\Nsigomegaks$ $\bz\to\omega\ks$ events
in the signal region
with a statistical significance ($\Sigma$) of 7.3,
where $\Sigma$ is defined as 
$\Sigma \equiv \sqrt{-2\ln(\lzero/\lnsig)}$, and
$\lzero$ and $\lnsig$ denote the maximum likelihoods
of the fits without and with the signal component, respectively.

%%%%%%%%%%%%%%%%%%%%%%%%%%%%%%%%%
\subsection{\boldmath $\bz\to\ks\piz$}
%%%%%%%%%%%%%%%%%%%%%%%%%%%%%%%%%
Candidate $\ks \to \pip\pim$ decays
are selected with the same criteria as those used for
the $\bz\to\phi\ks$ decay, except that we impose
a more stringent invariant mass requirement;
only pairs of oppositely charged pions that have an invariant mass
within 0.015 GeV/$c^2$
of the nominal $\ks$ mass are used.
The $\piz$ selection criteria are the same as 
those used for the $\bz\to\eta'\ks$ decay.

The $B$ meson signal region is defined as 
$-0.15$ GeV $< \dE <0.1$ GeV
and $5.27~{\rm GeV}/c^2 <\mb<5.29~{\rm GeV}/c^2$.
The $\dE$ distribution for $\ks\piz$ has a tail toward lower $\dE$.
The $\dE$ resolution is 0.047~GeV for the main component.
%due to $\gamma$ energy leakage in ECL. 
The width of the tail is about 0.1~GeV.
%The $\dE$ resolution is 54~MeV.
The dominant background is from continuum.
In addition, according to MC simulation, there is a small ($\sim 2\%$)
contamination from other charmless rare $B$ decays.
We use extended modified Fox-Wolfram moments,
which were applied for the selection of
the $\bz\to\piz\piz$ decay~\cite{Abe:2003yy},
to form a Fisher discriminant $\calf$.
We then combine likelihoods for $\calf$ and $\cos\theta_B$
to obtain the event likelihood ratio $\rsigbkg$ for
continuum suppression.

%The effect of possible $CP$ asymmetry of the $B$ decay background is 
%treated as a source of systematic error.
%from $B\overline{B}$ background events, which are dominated
%by other charmless rare $B$ decays.

As described below, we include events that do not have
$B$ decay vertex information in our fit
to obtain better sensitivity for
the $CP$-violation parameter $\cala$.
For events with vertex information,
the high-$\rsigbkg$ region is defined as
$\rsigbkg > 0.78$
and the low-$\rsigbkg$ region as
$0.4 < \rsigbkg \le  0.78$.
For events without vertex information,
the high-$\rsigbkg$ region is defined as
$\rsigbkg > 0.74~(0.76)$ for DS-I (DS-II),
and the low-$\rsigbkg$ region as
$0.4 < \rsigbkg \le  0.74~(0.76)$ for DS-I (DS-II).
%With the high-$rsigbkg$ requirement, 64\% of signal events remains and 
By the high-$\rsigbkg$ requirement, 95\% of continuum backgrounds
are rejected and 64\% of signal events remain.
As for the low-$\rsigbkg$ region, 84\% of continuum backgrounds are rejected 
and 26\% of signal events remain.

\begin{figure}
\includegraphics[width=0.32\textwidth]{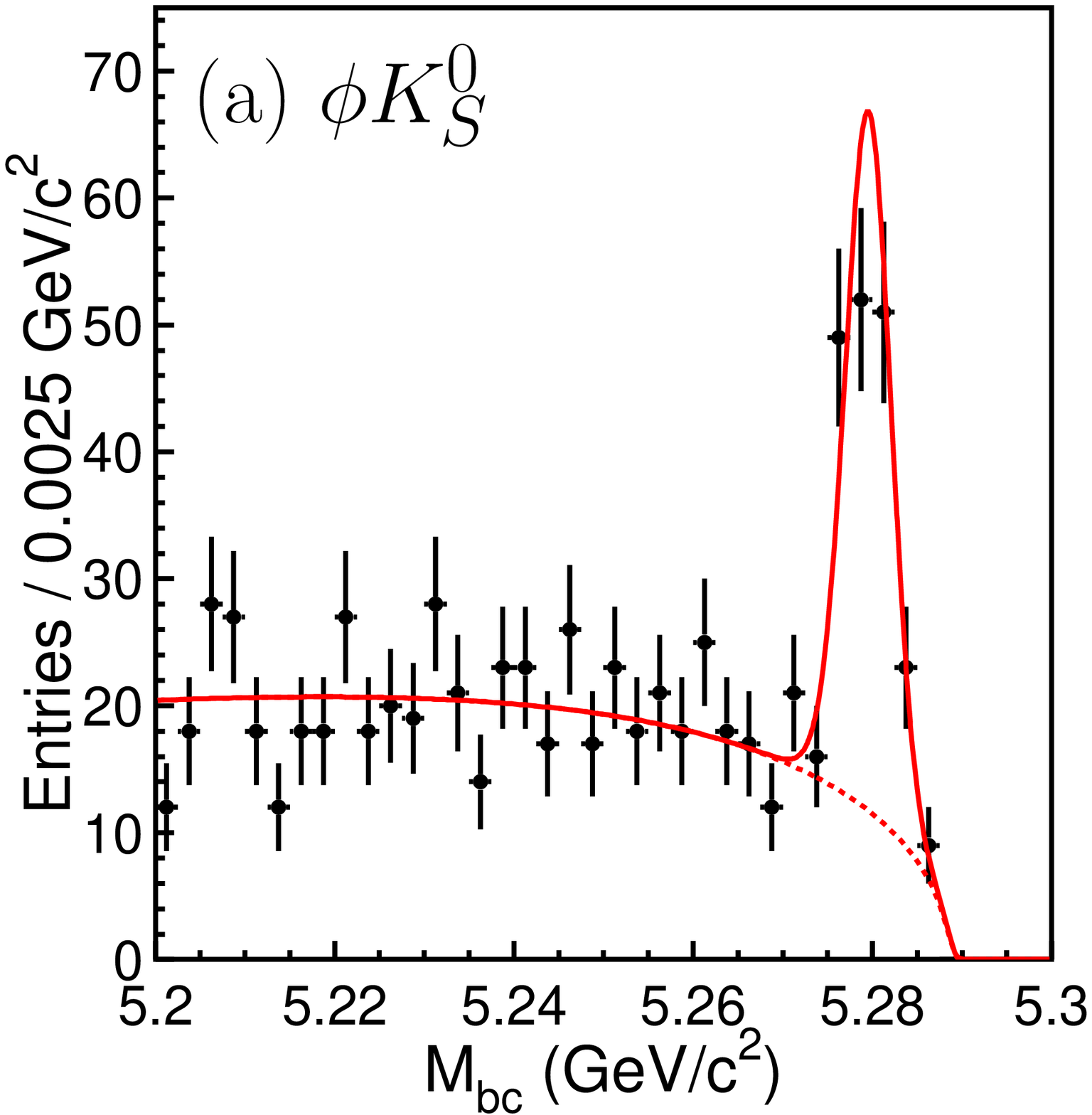}
\includegraphics[width=0.32\textwidth]{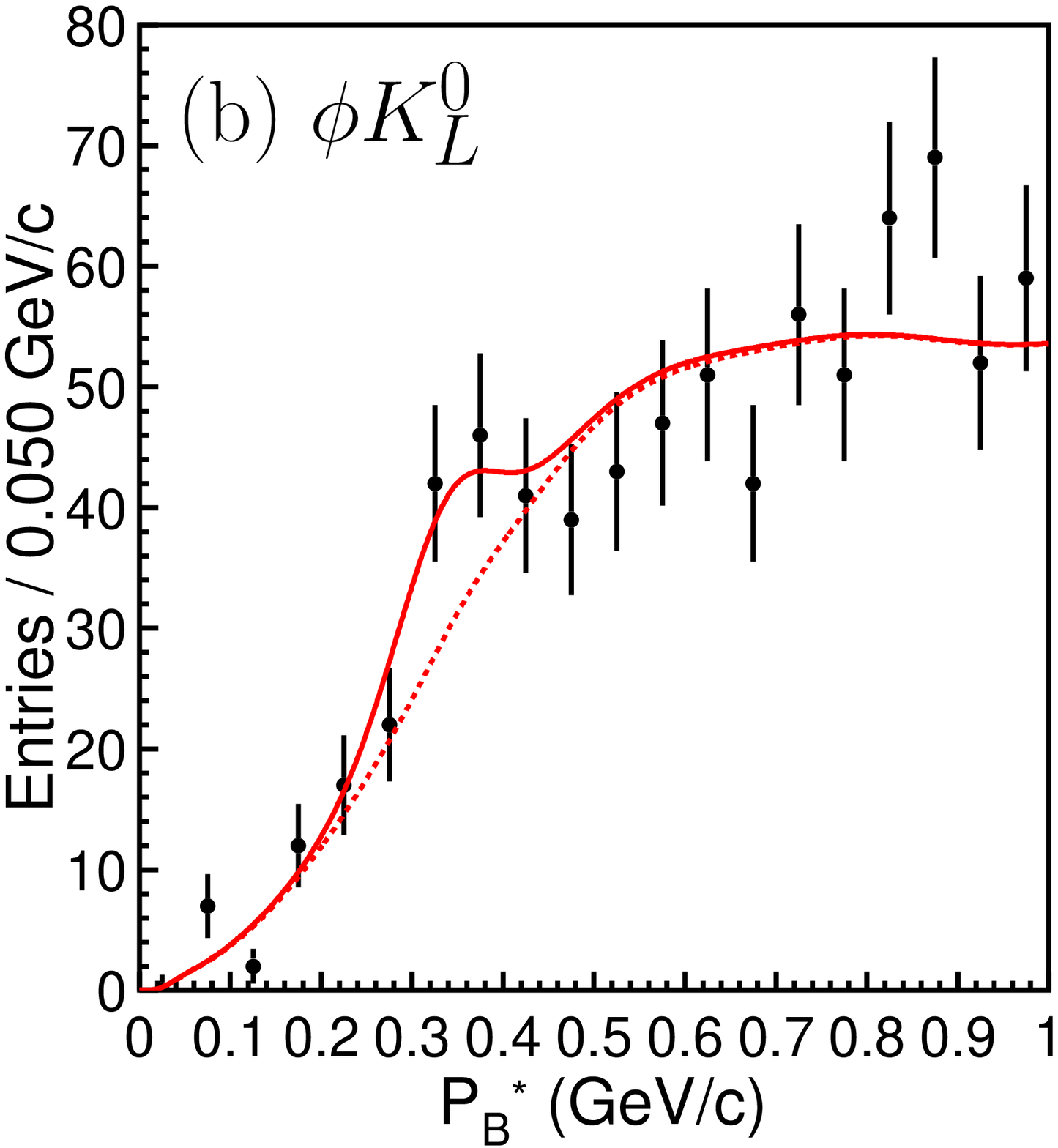}
\includegraphics[width=0.32\textwidth]{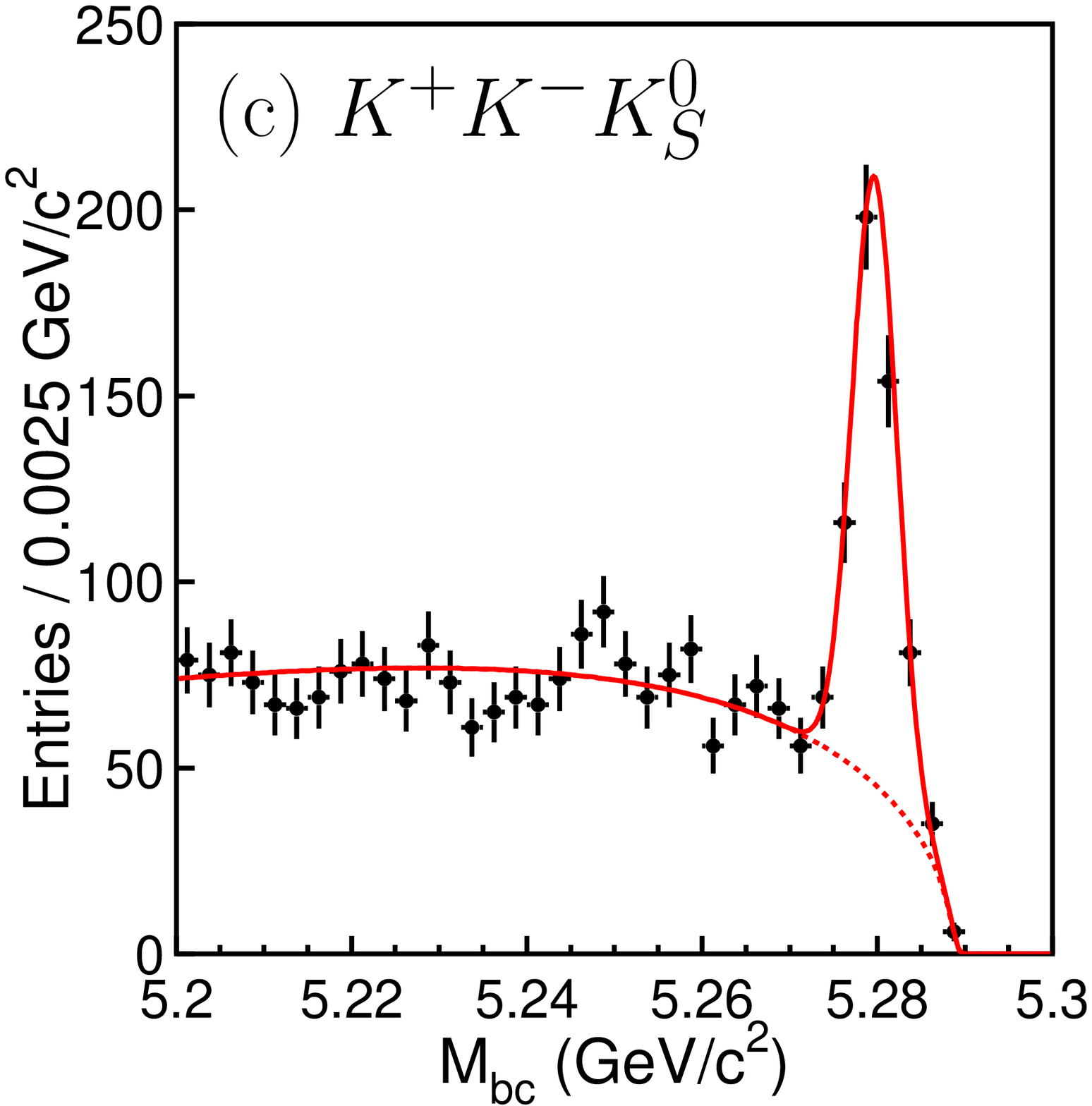}
\includegraphics[width=0.32\textwidth]{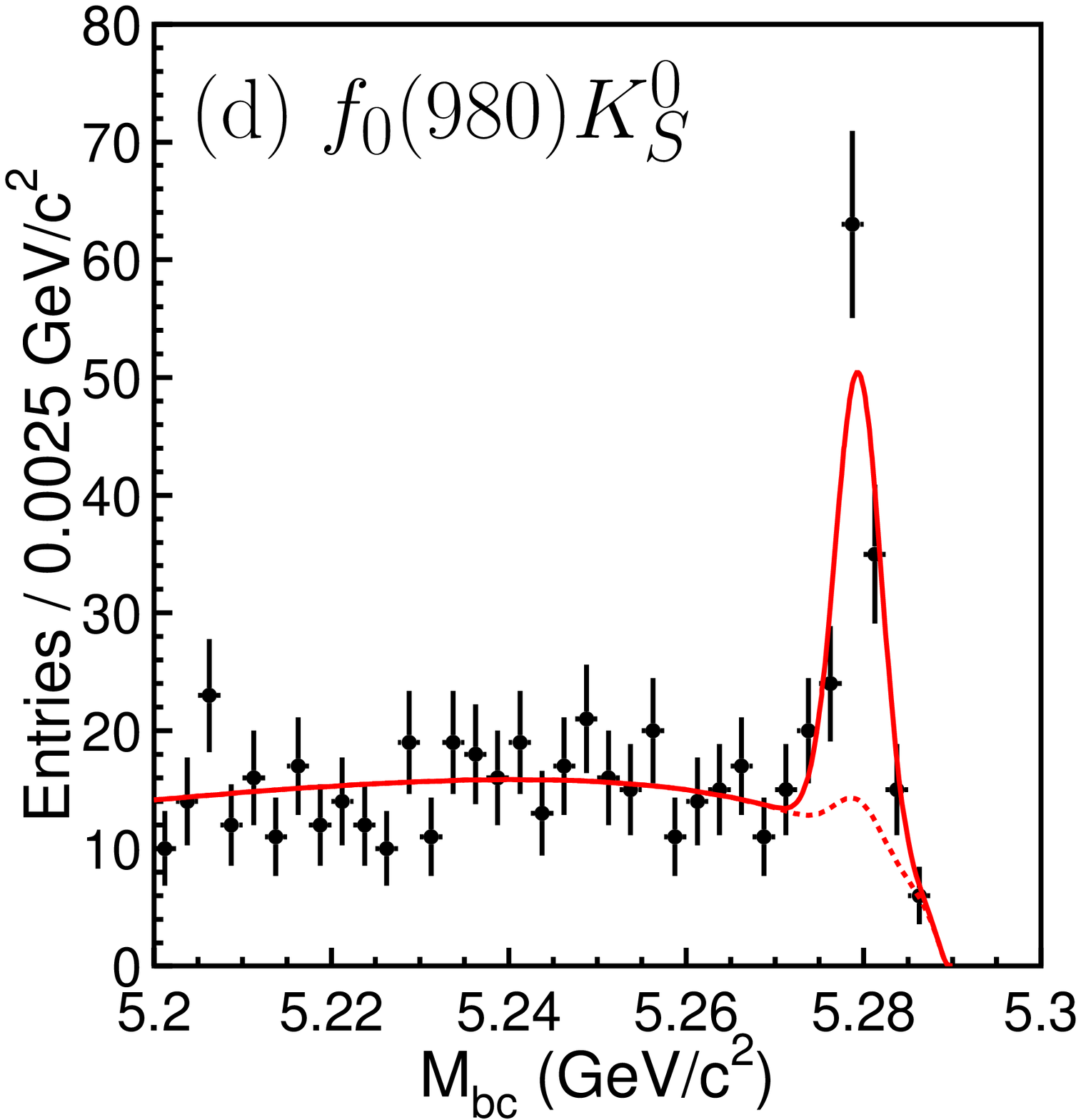}
\includegraphics[width=0.32\textwidth]{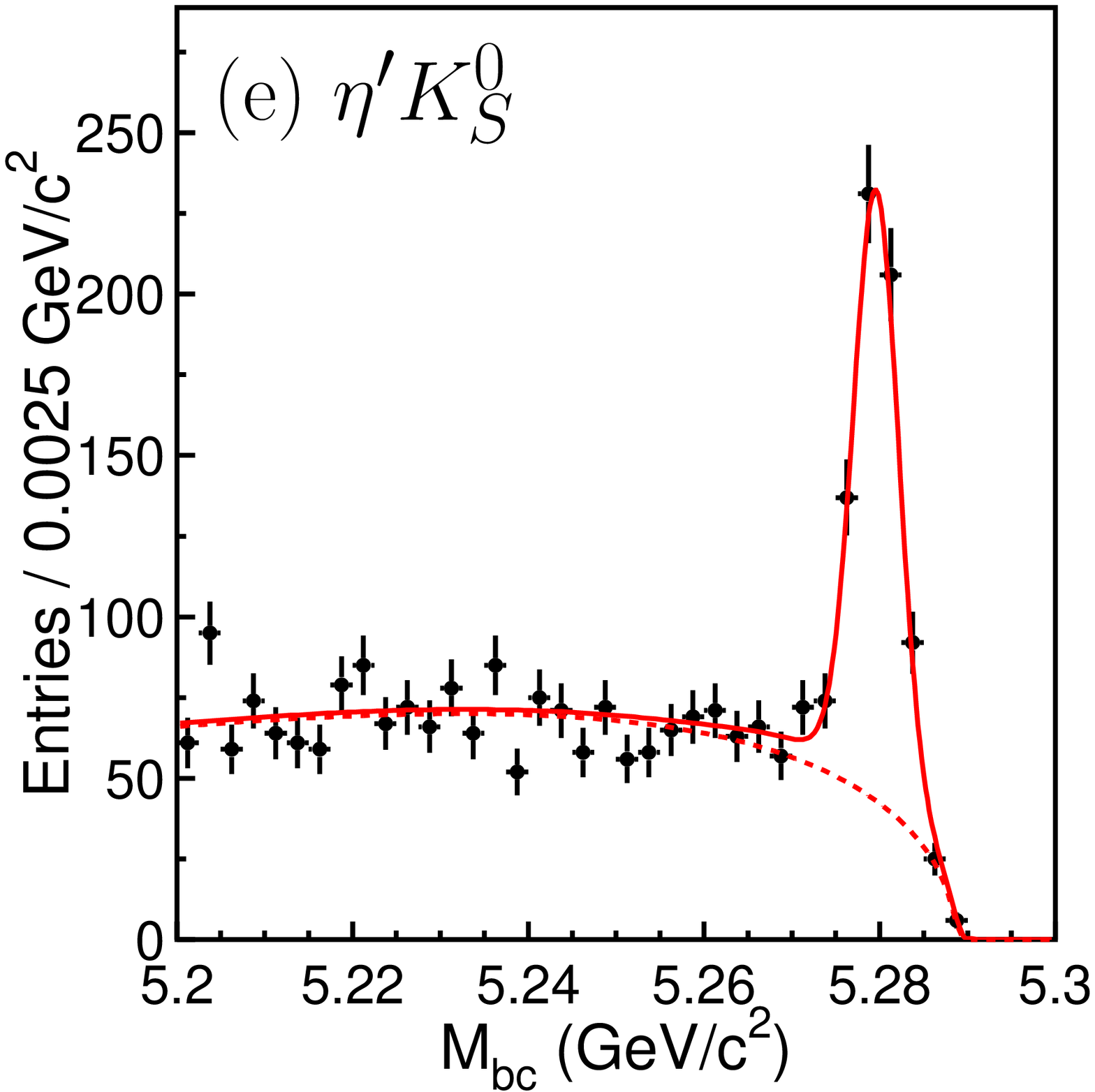}
\includegraphics[width=0.32\textwidth]{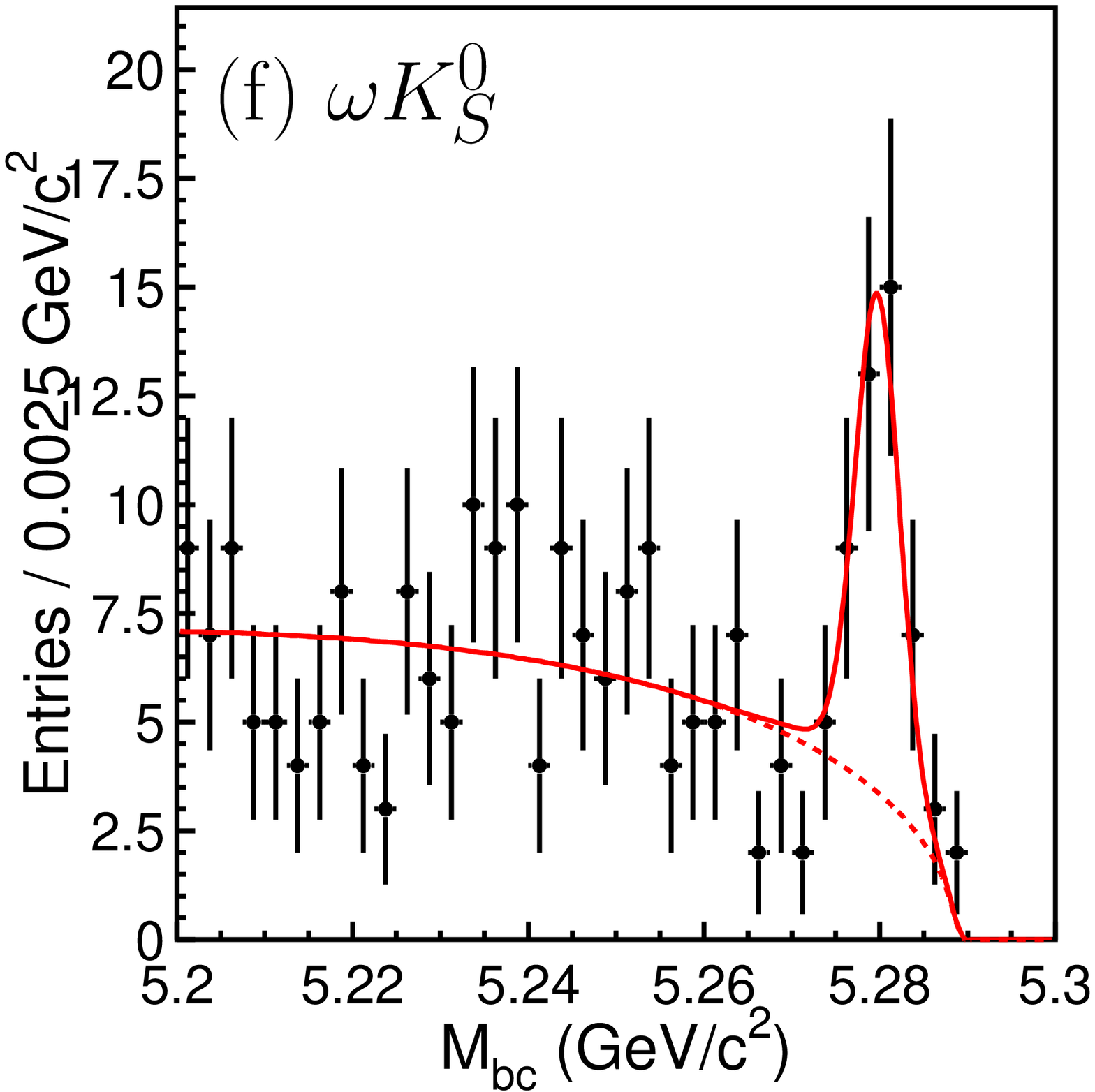}
\includegraphics[width=0.32\textwidth]{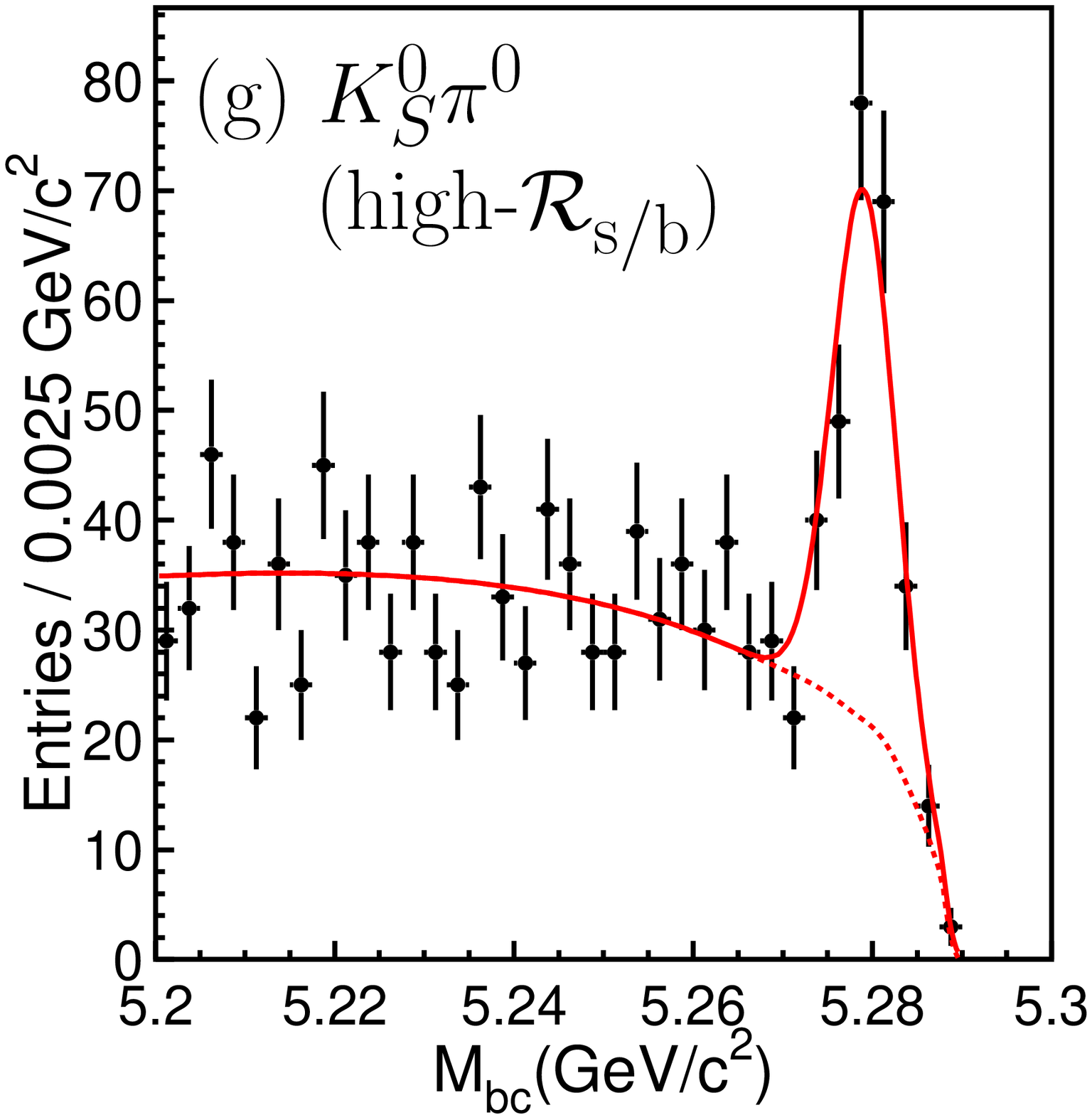}
\includegraphics[width=0.32\textwidth]{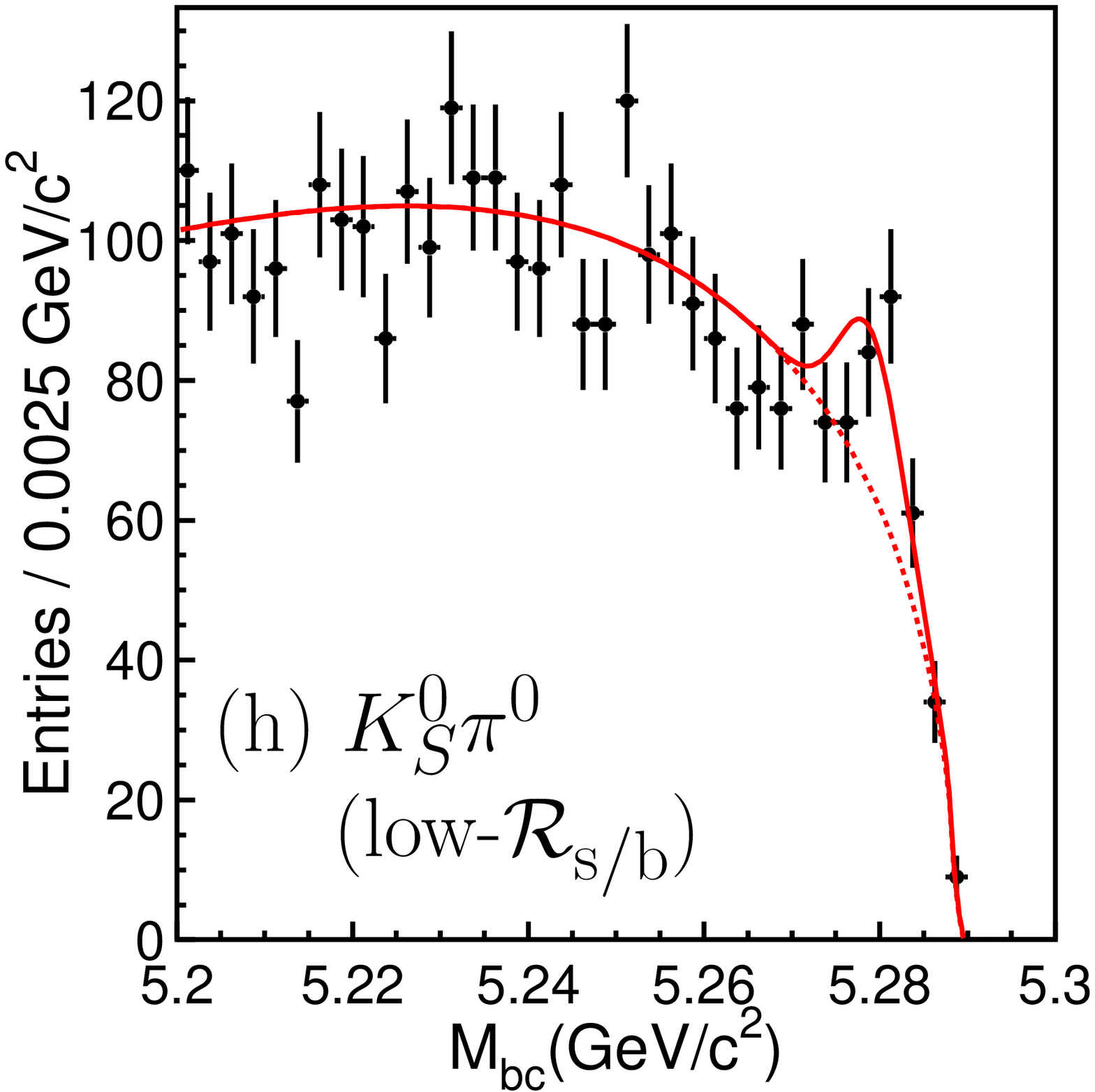}
\caption{$\mb$ distributions for
(a) $\bz\to\phi\ks$, (c) $\bz\to\kp\km\ks$, (d) $\bz\to\fzero\ks$,
(e) $\bz\to\eta'\ks$, (f) $\bz\to\omega\ks$, (g) $\bz\to\ks\piz$ 
(high-$\rsigbkg$), 
and (h) $\bz\to\ks\piz$ (low-$\rsigbkg$), 
within the $\dE$ signal region
and (b) $p_{B}^{\rm cms}$ distribution for $\bz\to\phi\kl$.
Solid curves show the fits to signal plus background distributions,
and dashed curves show the background contributions.
}\label{fig:mb}
%
%\rput[l](-6.5, 17.0)  {(a)~$\phi\ks$}
%\rput[l](-1.4, 17.0)  {(b)~$\phi\kl$}
%\rput[l]( 3.65,17.0)  {(c)~$\kp\km\ks$}
%\rput[l](-6.5, 12.0)  {(d)~$\fzero\ks$}
%\rput[l](-1.4, 12.0)  {(e)~$\eta'\ks$}
%\rput[l]( 3.65,12.0)  {(f)~$\omega\ks$}
%\rput[l](-4.0,  7.0)  {(g)~$\ks\piz$}
%\rput[l](-3.5,  6.5)   {(high-$\rsigbkg$)}
%\rput[l]( 1.0,  4.5){(h)~$\ks\piz$}
%\rput[l]( 1.5,  4.0)     {(low-$\rsigbkg$)}
%
\end{figure}

Figure~\ref{fig:mb}(g) shows
the $\mb$ distribution for the high-$\rsigbkg$ $\bz\to\ks\piz$ candidates
within the $\dE$ signal region
after flavor tagging and before vertex reconstruction.
Also shown in Fig.~\ref{fig:mb}(h) is the $\mb$ distribution
for the low-$\rsigbkg$ $\bz\to\ks\piz$ candidates.
The $\dE$ distributions for the high-$\rsigbkg$ and low-$\rsigbkg$ candidates 
are shown in Fig.~\ref{fig:de}(f) and (g), respectively.
The signal yield is determined
from an unbinned two-dimensional maximum-likelihood fit
to the $\dE$-$\mb$ distribution in the fit region defined as
$ 5.2~{\rm GeV/}c^2 < \mb < 5.29~{\rm GeV/}c^2$ 
and
$-0.5~{\rm GeV} < \dE < 0.5~{\rm GeV}$.
The $\bz\to\ks\piz$ signal distribution is modeled with
a Gaussian function for $\mb$ and with a
Crystal Ball function~\cite{cb-line} for $\dE$.
For the continuum background, 
we use the ARGUS parameterization for $\mb$
and a second-order Chebyshev function for $\dE$.
The $B$ decay background distribution is represented by
a smoothed histogram obtained from MC simulation. 
The fits yield $\NsigkspizH$ and $\NsigkspizL$ $\bz\to\ks\piz$ events
in the high-$\rsigbkg$ and 
low-$\rsigbkg$ signal regions, respectively, where the errors are
statistical only.
The same procedure after the vertex reconstruction yields
a total of $71\pm 13$ $\ks\piz$ events.

%%%%%%%%%%%%%%%%%%%%%%%%%
\subsection{Flavor Tagging}
\label{sec:flavor tagging}
%%%%%%%%%%%%%%%%%%%%%%%%%%
The $b$-flavor of the accompanying $B$ meson is identified
from inclusive properties of particles
that are not associated with the reconstructed $\bz \to \fCP$ 
decay. We use the same procedure that is used for the
$\sinbb$ measurement~\cite{bib:BELLE-CONF-0436}.
The algorithm for flavor tagging is described in detail
elsewhere~\cite{bib:fbtg_nim}.
We use two parameters, $\fq$ and $r$, to represent the tagging information.
%The first, $\fq$, is already defined in Eq.~(\ref{eq:psig}).
The first, $\fq$, is defined in Eq.~(\ref{eq:psig}).
The parameter $r$ is an event-by-event,
MC-determined flavor-tagging dilution factor
that ranges from $r=0$ for no flavor
discrimination to $r=1$ for unambiguous flavor assignment.
It is used only to sort data into six $r$ intervals listed in
Table~\ref{tab:wtag}.
The wrong tag fractions for the six $r$ intervals, 
$w_l~(l=1,6)$, and differences 
between $\bz$ and $\bzb$ decays, $\dwl$,
are determined from the data;
we use the same values
that were used for the $\sin 2\phi_1$ measurement~\cite{bib:BELLE-CONF-0436}
for DS-I.
Wrong tag fractions for DS-II are separately obtained 
with the same procedure 
and are listed in Table~\ref{tab:wtag}.
The total effective tagging efficiency for DS-II
is determined to be
$\eeff \equiv \sum_{l=1}^6 \epsilon_l(1-2w_l)^2 = \efftot$,
where 
$\epsilon_l$ is the event fraction for each $r$ interval
determined from the $\jpsi\ks$ data and is listed in
Table~\ref{tab:wtag}.
The error includes both statistical and systematic uncertainties.
We find that the wrong tag fractions for DS-II
are slightly smaller than those for DS-I. As a result,
the $\eeff$ value for DS-II is slightly larger than that for DS-I
($\eeff = \efftotdsone$).

%%%%%%%%%%%%%%%%%%%%%%%%%%%%%%%
\subsection{Vertex Reconstruction}
%%%%%%%%%%%%%%%%%%%%%%%%%%%%%%%
The vertex position for the $\fCP$ decay 
is reconstructed using charged tracks that have enough SVD hits;
at least one layer with hits on both sides 
and at least one additional $z$ hit in other layers for SVD-I,
and at least two layers with hits on both sides for SVD-II.
A constraint on the IP is also used with the selected tracks;
the IP profile is convolved with the finite $B$ flight length in the plane
perpendicular to the $z$ axis.
The pions from $\ks$ decays are not used
except in the analysis of $\bz\to\ks\piz$ decays.
The typical vertex reconstruction efficiency and $z$ resolution 
for $\bz\to\phi\ks$ decays
are 95\% and 78~$\mu$m, respectively.
Similar values are obtained for other $\fCP$ decays except for
$\bz\to\ks\piz$ decays.

The vertex
for $\bz\to\ks\piz$ decays is
reconstructed using
the $\ks$ trajectory and the IP constraint, where
both pions from the $\ks$ decay are required to
have enough SVD hits in the same way as that for other $\fCP$ decays.
The reconstruction efficiency depends both on
the $\ks$ momentum and on the SVD geometry;
the efficiency with SVD-II (32\%) is higher than
that with SVD-I (23\%)
because of the larger outer radius and the additional layer.
The typical $z$ resolution of the vertex reconstructed with the $\ks$ is
93~$\mu$m for SVD-I and 110~$\mu$m for SVD-II.

The $\ftag$ vertex determination with SVD-I
remains unchanged from the
previous publication~\cite{Abe:2003yt},
and is described in detail elsewhere~\cite{bib:resol_nim};
to minimize the effect of long-lived particles, 
secondary vertices from charmed hadrons and a small fraction of
poorly reconstructed tracks, we adopt an iterative procedure
in which the track that gives the largest contribution to the
vertex $\chi^2$ is removed at each step 
until a good $\chi^2$ is obtained.
The reconstruction efficiency was measured to be 93\%.
The typical $z$ resolution is $140\mu$m~\cite{bib:CP1_Belle}.

For SVD-II, we find that
the same vertex reconstruction algorithm results in
a larger outlier fraction when only
one track remains after the iteration procedure.
Therefore, in this case, we repeat the
iteration procedure with
a more stringent requirement on the SVD-II hit pattern;
at least two of the three outer layers have hits on both sides.
The resulting outlier fraction, which is described in
Sec.~\ref{sec:results}, is comparable to
that for SVD-I, while 
the inefficiency caused by this change is small (2.5\%).

%%%%%%%%%%%%%%%%%%%%%%%%%%%%%%
\subsection{Summary of Signal Yields}
%%%%%%%%%%%%%%%%%%%%%%%%%%%%%%
The signal yields for $\bz\to\fCP$ decays, $\nsig$,
after flavor tagging and vertex reconstruction
(before the vertex reconstruction for the
decay $\bz\to\ks\piz$)
are summarized in Table~\ref{tab:num}.
The signal purities are also listed in the table.

%%%%%%%%%%%%%%%%%%%%%%%%%%%%%%%%%%%%%%%%%%%%%%%%
\section{Results of {\boldmath $CP$} Asymmetry Measurements}
\label{sec:results}
%%%%%%%%%%%%%%%%%%%%%%%%%%%%%%%%%%%%%%%%%%%%%%%%
We determine $\cals$ and $\cala$ for each mode by performing an unbinned
maximum-likelihood fit to the observed $\Dt$ distribution.
The probability density function (PDF) expected for the signal
distribution, ${\cal P}_{\rm sig}(\Dt;\cals,\cala,\fq,w_l,\dwl)$, 
is given by Eq.~(\ref{eq:psig}) incorporating
the effect of incorrect flavor assignment. The distribution is
convolved with the
proper-time interval resolution function 
$R_{\rm sig}(\Dt)$,
which takes into account the finite vertex resolution. 

For the decays $\bz\to\phi\ks$, $\kp\km\ks$, $\phi\kl$, $\fzero\ks$, $\eta'\ks$
and $\omega\ks$,
we use flavor-specific $B$ decays governed by
semileptonic or hadronic $b\to c$ transitions
to determine the resolution function.
We perform a simultaneous multiparameter fit to these high-statistics
control samples to obtain the resolution function parameters,
wrong-tag fractions (Section~\ref{sec:flavor tagging}), 
$\dmd$, $\taubp$ and $\taubz$.
We use the same resolution function used for
the $\sinbb$ measurement for DS-I~\cite{bib:BELLE-CONF-0436}.
For DS-II, the following modifications are introduced:
a sum of two Gaussian functions is used
to model the resolution of the $\fCP$ vertex
while a single Gaussian function is used for DS-I;
a sum of two Gaussian functions is used to model
the resolution of the tag-side vertex
obtained with one track and the IP constraint,
while a single Gaussian function is used for DS-I.
These modifications are needed to account for
differences between SVD-I and SVD-II, as well as
different background conditions in DS-I and DS-II.
We test the new resolution parameterization using
MC events on which we overlay
beam-related background taken from data.
A fit to the MC sample yields correct values
for all parameters. 
With the multiparameter fit to data, we find that
the standard deviation of the main Gaussian component
of the resolution function
is reduced from 78 $\mu$m to 55 $\mu$m, which
is consistent with our expectation from
the improved impact parameter resolution of SVD-II~\cite{Ushiroda}.
The same fit also yields
$\taubz = 1.519\pm 0.010$ ps,
$\taubp = 1.652\pm 0.012$ ps
and 
$\dmd = 0.516\pm 0.007~{\rm ps}^{-1}$, where errors are statistical only.
The results are consistent with those obtained with 
DS-I~\cite{bib:BELLE-CONF-0436}
and also with the world average values~\cite{bib:PDG2004}.
Thus we conclude that the resolution of SVD-II is well understood.

For the decay $\bz\to\ks\piz$,
we use the resolution function described above
with additional parameters that rescale vertex
errors. The rescaling parameters depend on
the detector configuration (SVD-I or SVD-II), 
SVD hit patterns of charged pions from the $\ks$ decay,
and $\ks$ decay vertex position in the plane 
perpendicular to the beam axis.
These parameters are determined from a fit to the 
$\Dt$ distribution of $\bz\to\jpsi\ks$ data.
Here the $\ks$ and the IP constraint are used for the
vertex reconstruction, the $\bz$ lifetime is
fixed at the world average value, and $b$-flavor tagging
information is not used so that the 
expected PDF is
an exponential function convolved with
the resolution function.

We check the resulting resolution function
by also reconstructing the vertex with
leptons from $\jpsi$ decays and the IP constraint.
We find that the distribution of the
distance between the vertex positions obtained with
the two methods is well represented by
the obtained resolution function convolved with
the well-known resolution for the $\jpsi$ vertex.
Finally, we also perform a fit to the $\bz\to\jpsi\ks$ sample
with $b$-flavor information and obtain
$\cals_{\jpsi\ks} = +0.68\pm 0.10$(stat) and
$\cala_{\jpsi\ks} = +0.02\pm 0.04$(stat), which
are in good agreement with the world average values.
Thus, we conclude that 
the vertex resolution for the $\bz\to\ks\piz$ decay
is well understood.

We determine the following likelihood value for each
event:
\begin{eqnarray}
P_i
&=& (1-\fol)\int \biggl[
\fsig{\cal P}_{\rm sig}(\Dt')R_{\rm sig}(\Dt_i-\Dt') \nonumber \\
&+&(1-\fsig){\cal P}_{\rm bkg}(\Dt')R_{\rm bkg}(\Dt_i-\Dt')\biggr]
d(\Dt')  \nonumber \\
&+&\fol P_{\rm ol}(\Dt_i),
\label{eq:likelihood}
\end{eqnarray}
where $P_{\rm ol}(\Dt)$ is a broad Gaussian function that represents
an outlier component with a small fraction $\fol$~\cite{bib:BELLE-CONF-0436}.
The width of the outlier component for DS-II is determined to be
$(44\pm 5)$~ps; the fractions of the outlier components are
$(3.1\pm1.2)\times 10^{-4}$ for events with the $\ftag$ 
vertex reconstructed with more than one track,
and $(1.2\pm 0.1)\times 10^{-2}$ for the case only one track is used.
These values are comparable to those for DS-I~\cite{bib:BELLE-CONF-0436}.
The signal probability $\fsig$ depends on the $r$ region and
is calculated on an event-by-event basis
as a function of $\pbstar$ for the $\bz\to\phi\kl$ decay and
as a function of $\dE$ and $\mb$ for the other modes.
A PDF for background events, ${\cal P}_{\rm bkg}(\Dt)$,
is modeled as a sum of exponential and prompt components, and
is convolved with a sum of two Gaussians $R_{\rm bkg}$.
Parameters in ${\cal P}_{\rm bkg} (\Dt)$ and $R_{\rm bkg}$ 
for continuum background are determined by the fit to the $\Dt$ distribution
for events outside the $\dE$-$\mb$ signal region except for the 
$\bz\to\phi\kl$ decay.
For the $\bz\to\phi\kl$ decay, we use $\pbstar$ sideband events to obtain the
parameters.
Parameters in ${\cal P}_{\rm bkg}(\Dt)$ and $R_{\rm bkg}$
for $B\overline{B}$ background events in $\bz\to\eta'\ks$,
$\bz\to\ks\piz$ and $\bz\to\phi\kl$ decays are determined from MC simulation. 

%of a background-enhanced control sample~\cite{bib:BBbg};
%i.e. events outside of the $\dE$-$\mb$ signal region.
We fix $\tau_\bz$ and $\dmd$ at
their world average values~\cite{bib:PDG2004}.
We assume no $CP$ asymmetry in the background $\Delta t$ distributions and 
possible $CP$ asymmetries in the $B$ decay backgrounds are treated as
sources of systematic error.
In order to reduce the statistical error on $\cala$,
we include events without vertex information
in the analysis of $\bz\to\ks\piz$.
The likelihood value in this case is obtained by integrating 
Eq.~(\ref{eq:likelihood}) over $\Dt_i$.

\begin{figure}
\includegraphics[width=0.32\textwidth]{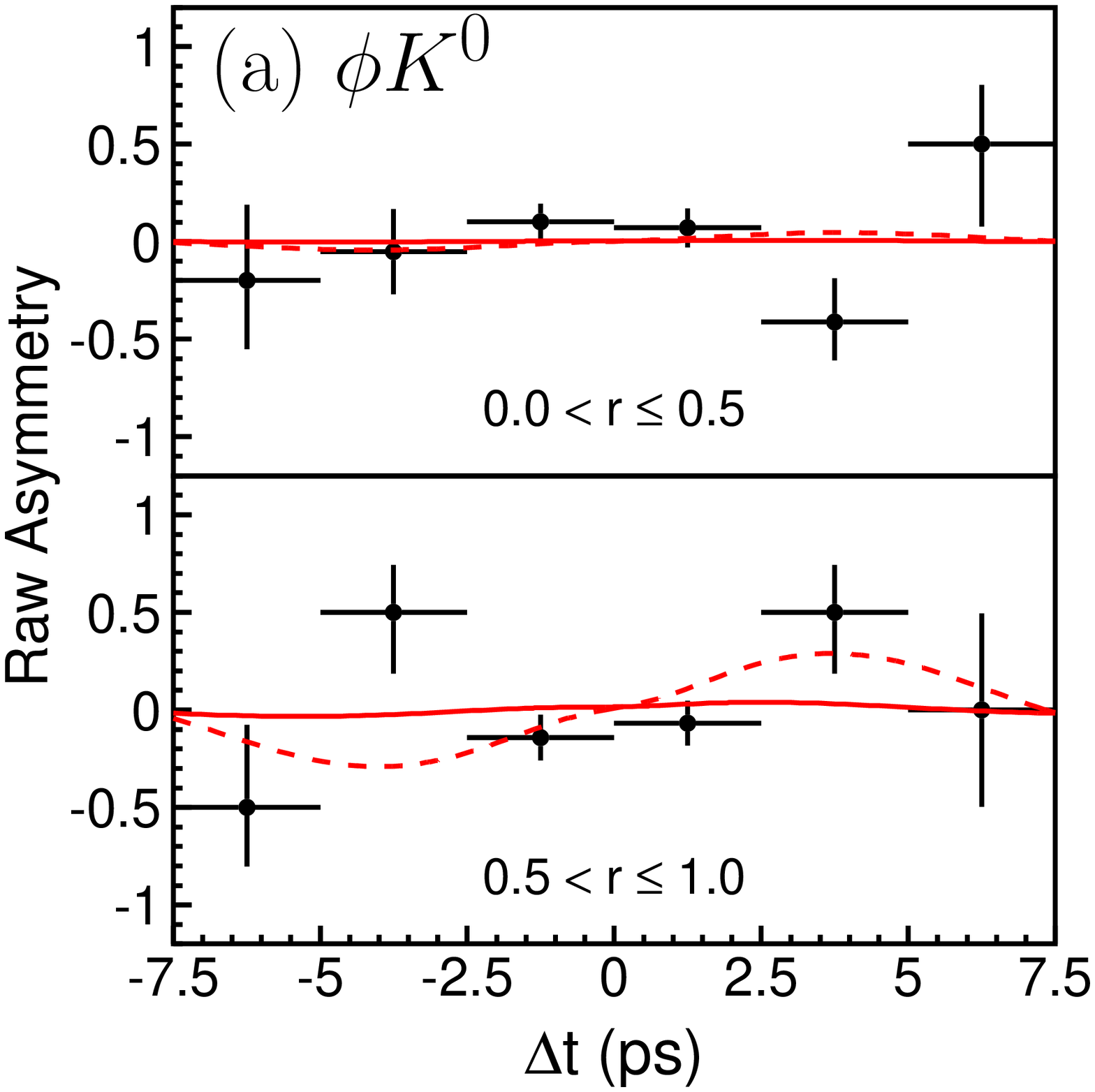}
\includegraphics[width=0.32\textwidth]{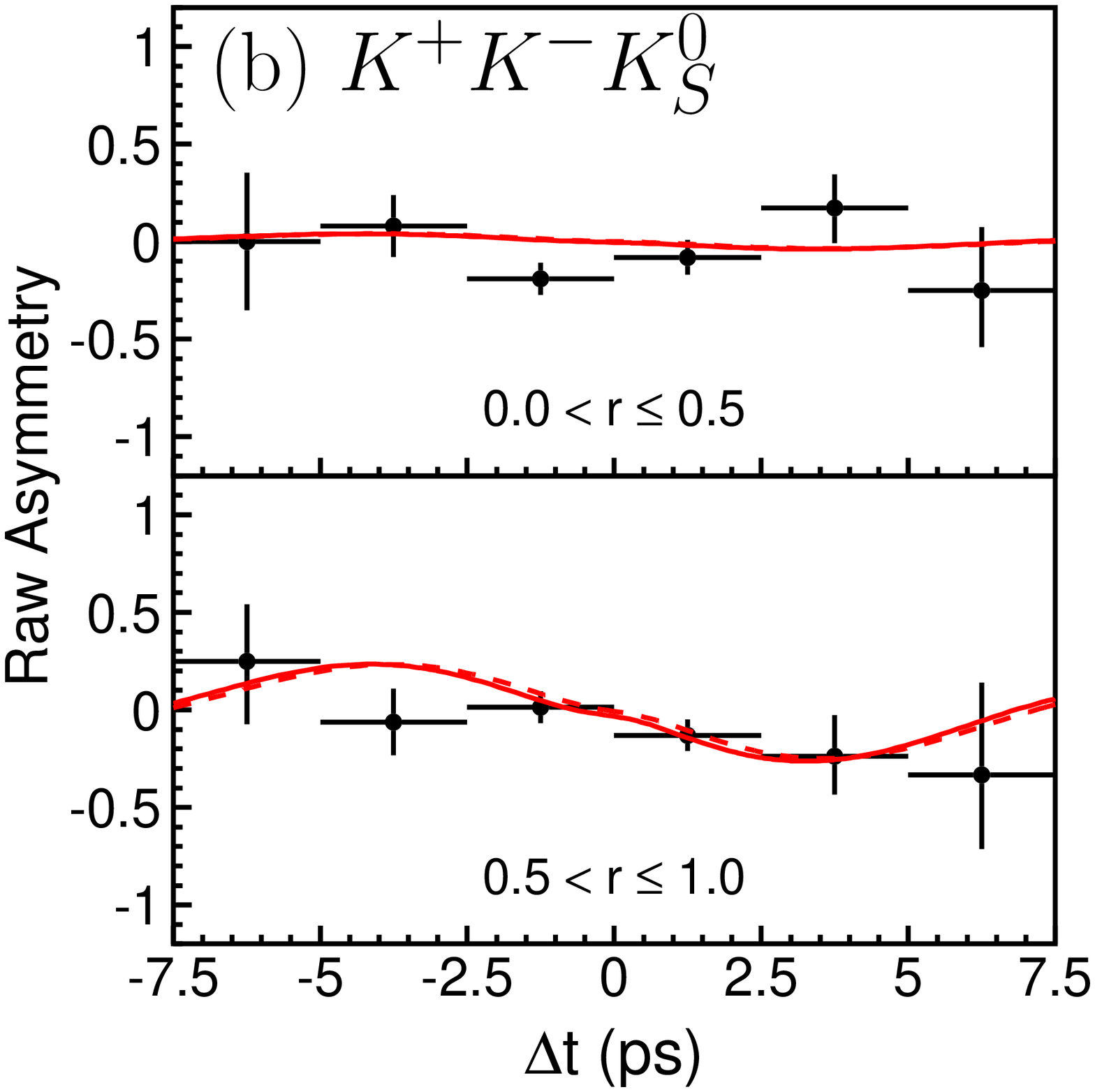}
\includegraphics[width=0.32\textwidth]{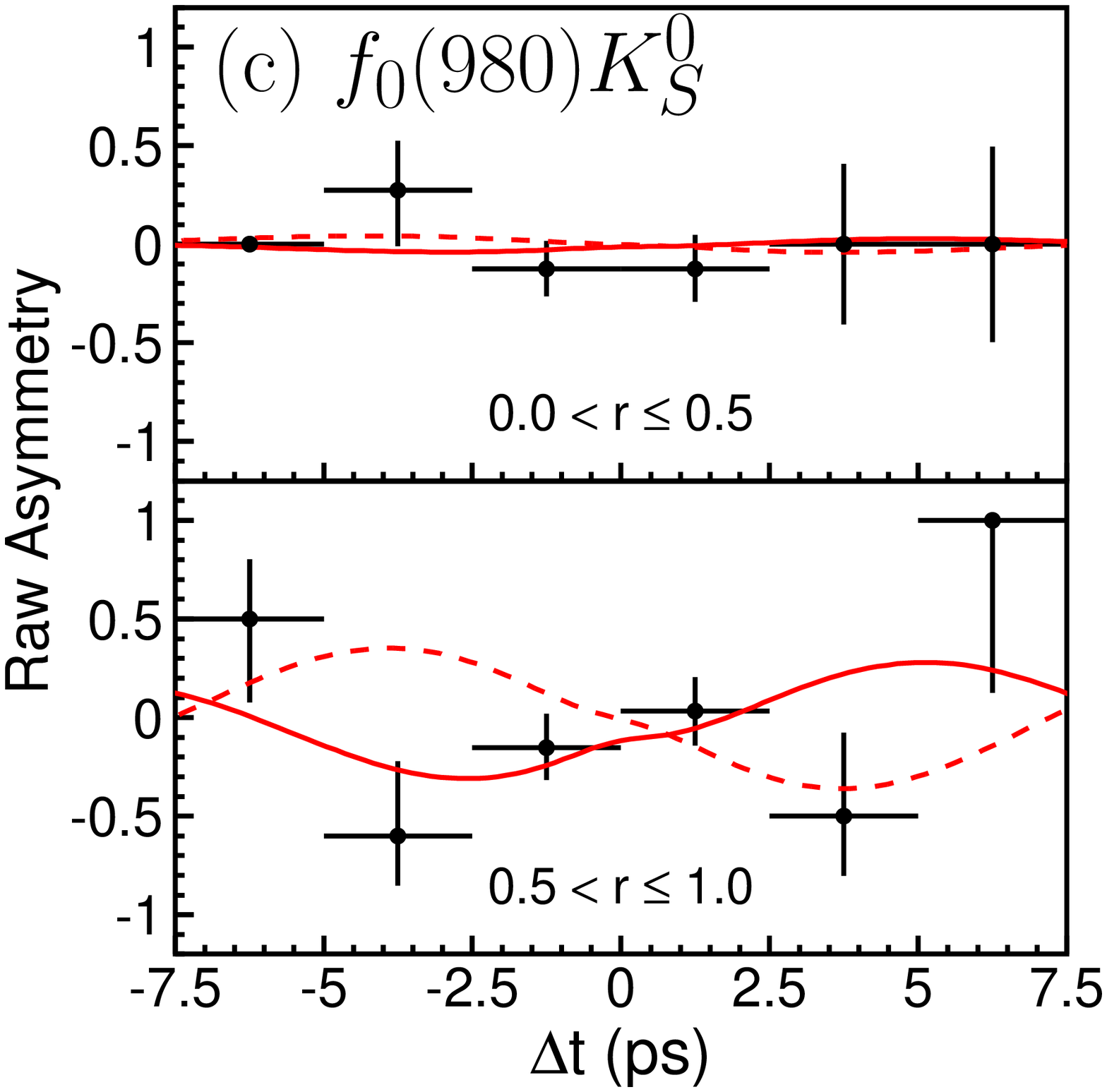}
\includegraphics[width=0.32\textwidth]{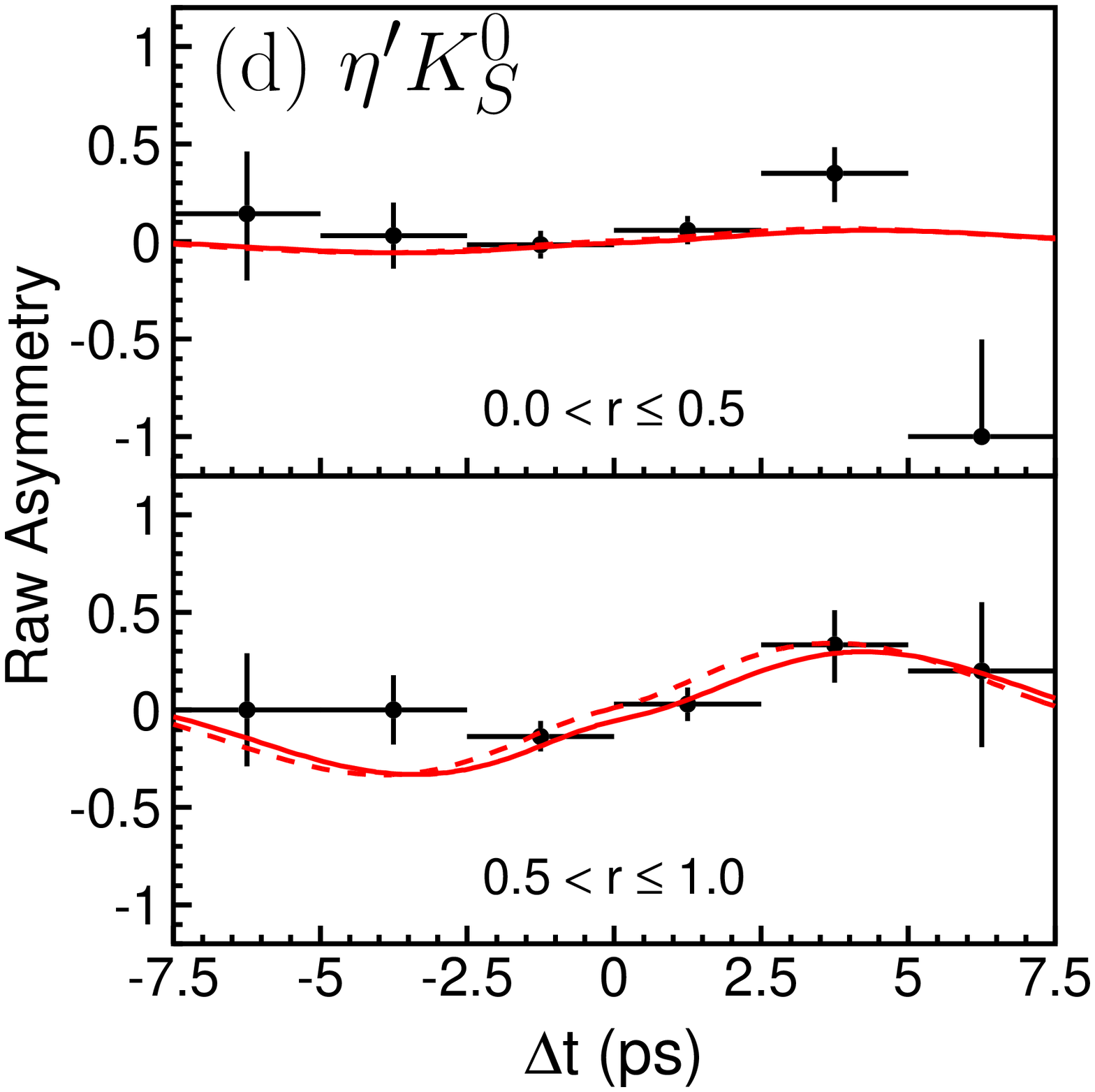}
\includegraphics[width=0.32\textwidth]{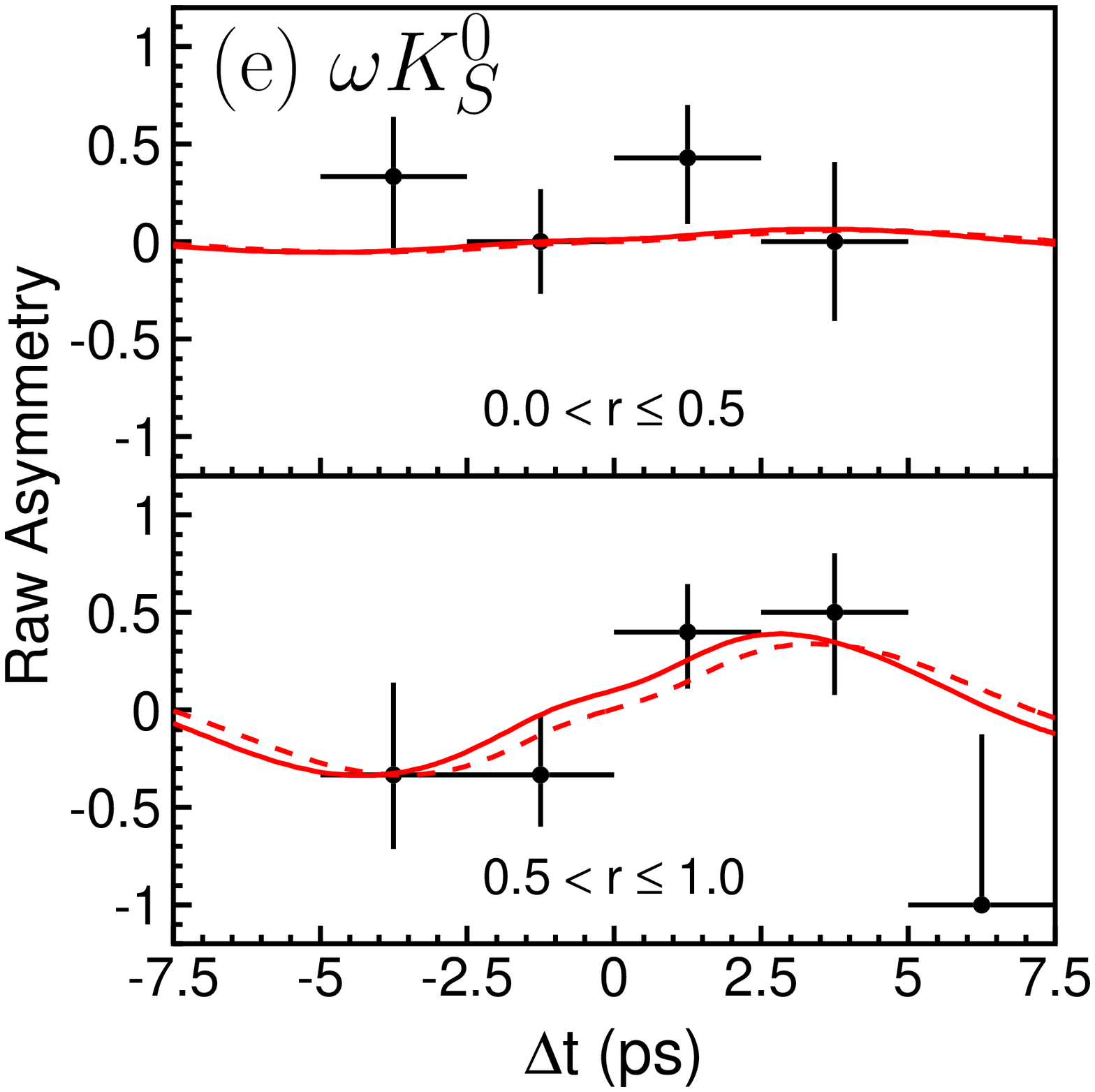}
\includegraphics[width=0.32\textwidth]{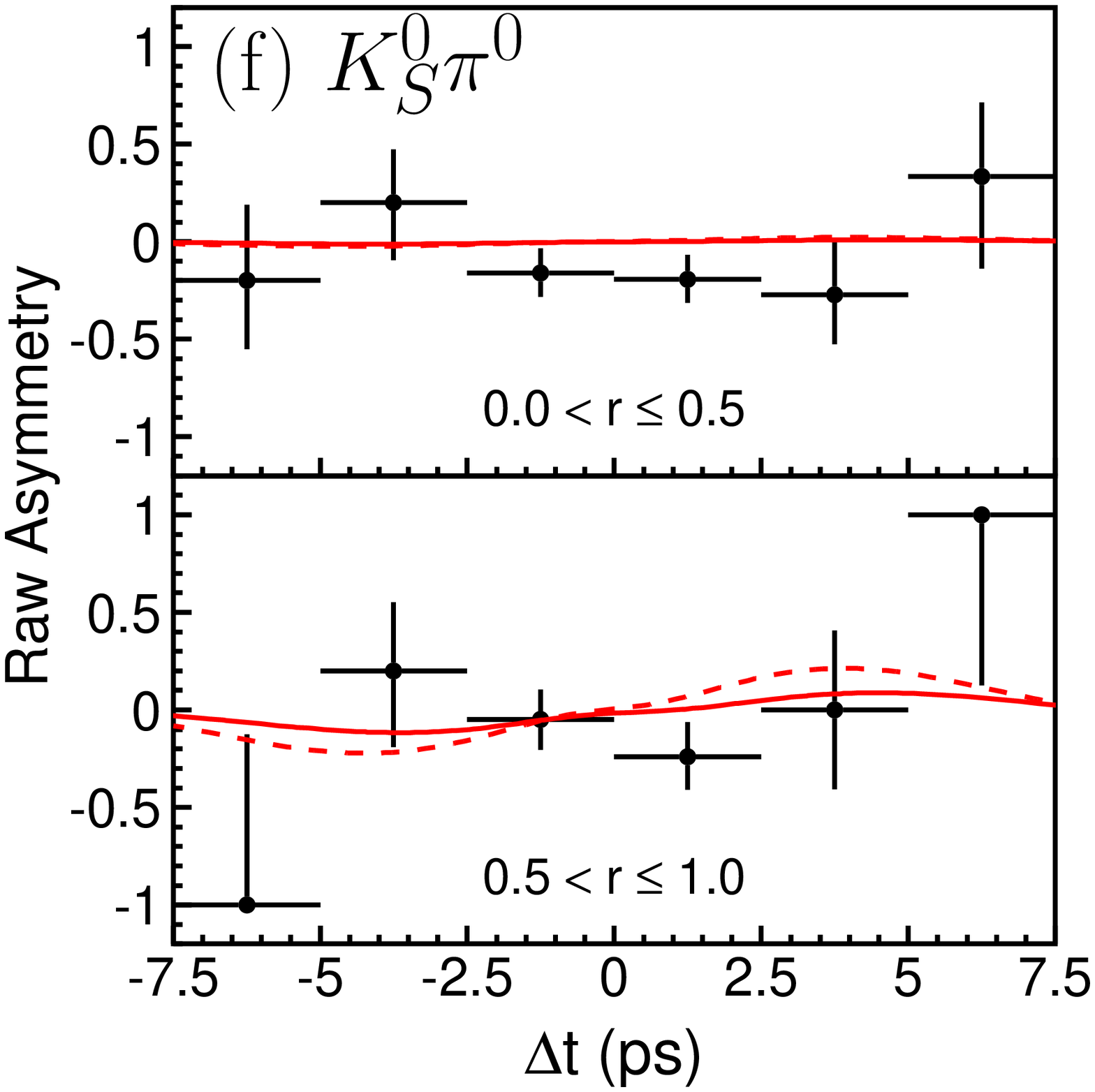}
\caption{
Raw asymmetry in each $\Dt$ bin with $0 < r \le 0.5$ (top)
and with $0.5 < r \le 1.0$ (bottom) for 
(a) $\bz\to\phi\kz$, 
(b) $\bz\to\kp\km\ks$, (c) $\bz\to\fzero\ks$,
(d) $\bz\to\eta'\ks$, (e) $\bz\to\omega\ks$, 
and (f) $\bz\to\ks\piz$.
The solid curves show the results of the 
unbinned maximum-likelihood fits.
The dashed curves show the SM expectation with $\sin2\phi_1$ = +0.73 
and $\cala$ = 0.}
%
%\rput[l](-6.9, 12.7)  {(a)~$\phi\kz$}
%\rput[l](-1.5, 12.7)  {(b)~$\kp\km\ks$}
%\rput[l]( 3.85,12.7)  {(c)~$\fzero\ks$}
%\rput[l](-6.9,  7.4)  {(d)~$\eta'\ks$}
%\rput[l](-1.5,  7.4)  {(e)~$\omega\ks$}
%\rput[l]( 3.75, 7.4)  {(f)~$\ks\piz$}
%
\label{fig:asym}
\end{figure}

The only free parameters in the final fit
are $\cals$ and $\cala$, which are determined by maximizing the
likelihood function
$L = \prod_iP_i(\Dt_i;\cals,\cala)$
where the product is over all events.
Table~\ref{tab:result} summarizes
the fit results of $\cals$ and $\cala$.
We define the raw asymmetry in each $\Dt$ bin by
$(N_{q=+1}-N_{q=-1})/(N_{q=+1}+N_{q=-1})$,
where $N_{q=+1(-1)}$ is the number of 
observed candidates with $q=+1(-1)$~\cite{footnote:phikz}.
Figures~\ref{fig:asym}(a-f) show the raw asymmetries in two regions of the flavor-tagging
parameter $r$. While the numbers of signal events in the two regions are 
similar,
the effective tagging efficiency is much larger 
and the background dilution is smaller in the region $0.5 < r \le 1.0$.
Note that these projections onto the $\Delta t$ axis do not take into
account event-by-event information (such as the signal fraction, the
wrong tag fraction and the vertex resolution), which is used in the
unbinned maximum-likelihood fit.

%%%%%%%%%%%%%%%%%%%%%%%%%%%
%\subsection{Systematic Error}
%%%%%%%%%%%%%%%%%%%%%%%%%%%
Tables~\ref{tab:ssyserr} and \ref{tab:asyserr}
list the systematic errors on $\cals$ and $\cala$, respectively.
The total systematic errors are obtained
by adding each contribution in quadrature,
and are much smaller than the statistical errors for all modes.

To determine the systematic error that arises from
uncertainties in the vertex reconstruction,
the track and vertex selection criteria
are varied to search for possible systematic biases.
Small biases in the $\Dz$ measurement 
are observed in $e^+e^-\to\mu^+\mu^-$ and other control
samples. Systematic errors 
are estimated by applying special correction functions
to account for the observed biases, repeating
the fit, and comparing the obtained values with the nominal results.
The systematic error due to the IP constraint 
in the vertex reconstruction is estimated by
varying ($\pm10 \mu$m) the smearing used to account for the
$B$ flight length.
Systematic errors due to imperfect SVD alignment
are determined
from MC samples that have artificial misalignment effects
to reproduce impact-parameter resolutions observed in data.

Systematic errors due to uncertainties in the wrong tag
fractions are studied by varying
the wrong tag fraction individually for each $r$ region.
Systematic errors due to uncertainties in the resolution function
are also estimated by varying each resolution parameter obtained from
data (MC) by $\pm 1\sigma$ ($\pm 2\sigma$), repeating the fit
and adding each variation in quadrature.
Each physics parameter such as $\taubz$ and $\dmd$
is also varied by its error.
A possible fit bias is examined by fitting a large number of MC events.

Systematic errors from uncertainties in the background fractions
and in the background $\Dt$ shape
are estimated by varying each background parameter obtained
from data (MC) by $\pm 1\sigma$ ($\pm 2\sigma$).

Additional sources of systematic errors are 
considered for $B$ decay backgrounds
that are neglected in the PDF.
We consider uncertainties both in their fractions
and $CP$ asymmetries; for modes that have
non-vanishing $CP$ asymmetries, we conservatively
vary the $CP$-violation parameters within the
physical region and take the largest variation
as the systematic error.
The effect of backgrounds from $\kp\km\ks$ and 
$\fzero\ks~(\fzero\to\kp\km)$
in the $\bz\to\phi\ks$ sample is considered.
Uncertainties from
$B\to\phi K^*$ and other rare $B$ decay backgrounds
in the $\bz\to\phi\kl$ sample
are also taken into account.
Effects of possible $CP$ asymmetries in $B$ decay backgrounds for $\ks\piz$ 
and $\fzero\ks$ are evaluated.
%For $\bz\to\kp\km\ks$, the effect of the $\kp\km$ from $D^0$ decays
%is evaluated by repeating the fit excluding $\kp\km$ pairs with an invariant
%mass within 10~MeV$/c^2$ of the nominal $D^0$ mass.
The peaking background fraction in the $\bz\to\fzero\ks$ sample
depends on the functions used in the fit to the 
$\pip\pim$ invariant mass distribution.
The systematic errors due to the uncertainties of the masses and widths of
the resonances used in the fit are estimated.
The width of $\fzero$ is varied between 0.04~GeV$/c^2$ and 0.1~GeV$/c^2$.
The mass and width of the $\fx$
are varied by $\pm0.03$~GeV$/c^2$ and $\pm0.1$~GeV$/c^2$, respectively.
The effect of possible interference between resonant and non-resonant
amplitudes, which is neglected in the nominal analysis, is also evaluated.
We perform a fit to the $\pip\pim$ distribution
of a MC sample generated with
interfering amplitudes and phases for $B\to K\pi\pi$ decays measured
from data~\cite{Garmash:2003er}. The observed difference in the
signal yield from the true value is taken into account in the 
systematic error determination.
We also repeat the fit to the $\Dt$ distribution
ignoring the contribution of the peaking background.
The differences in $\cals$ and $\cala$ from
our nominal results are included in the systematic error.

Finally, we investigate the effects of interference between
CKM-favored and CKM-suppressed $B\to D$ transitions in
the $\ftag$ final state~\cite{Long:2003wq}.
A small correction to the PDF for the signal distribution
arises from the interference.
We estimate the size of the correction using the $\bzdslnu$ 
sample. We then generate MC pseudoexperiments
and make an ensemble test to obtain systematic biases
in $\cals$ and $\cala$. 
In general, we find effects
on $\cals$ are negligibly small, while
there are sizable possible shifts in $\cala$.

%%%%%%%%%%%%%%%%%%%%%%
%\section{Cross Check}
%%%%%%%%%%%%%%%%%%%%%%
Various crosschecks of the measurements are performed.
We reconstruct charged $B$ meson decays
that are the counterparts of the $\bz\to\fCP$ decays
and apply the same fit procedure.
All results for the $\cals$ term are consistent with no 
$CP$ asymmetry, as expected. 
Lifetime measurements are also performed for 
the $\fCP$ modes and the corresponding charged $B$ decay modes.
The fits yield
$\taubz$ and $\taubp$ values consistent with the world average values.
MC pseudoexperiments are generated for each decay mode to
perform ensemble tests.
We find that the statistical errors obtained
in our measurements are all consistent
with the expectations from the ensemble tests.

For the $\bz\to\phi\kz$ decay,
a fit to DS-I alone yields
$\cals = -0.68\pm 0.46$(stat) and 
$\cala =  0.00 \pm 0.28$(stat),
while a fit to DS-II alone yields
$\cals = +0.80\pm 0.45$(stat) and $\cala = +0.15 \pm 0.33$(stat).
Note that the results for DS-I
differ from our previously published results
$\cals = \SphiksResPrv$ and $\cala = \AphiksResPrv$~\cite{Abe:2003yt},
as the decays $\bz\to\phi\kl$ and $\phi\ks~(\ks\to\piz\piz)$ are
included in this analysis.
%~\cite{footnote:svd1_phiks}.
Fit results to $\bz\to\phi\ks~(\ks\to\pip\pim)$ decays in DS-I 
in this analysis are $\cals = -0.97\pm0.50$(stat) and $\cala =
0.15\pm0.29$(stat), which are consistent with the previous results.
From MC pseudoexperiments, 
the probability that the difference between the $\cals$ values in
DS-I and DS-II is larger than the observed difference (1.46)
is estimated to be 4.5\%.
Fits to only $\phi\ks$ decays and only $\phi\kl$ decays
yield $\cals = \SphiksResultStat$ and $\cala = \AphiksResultStat$,
and $-\cals = \mSphiklResultStat$ and $\cala = \AphiklResultStat$, 
respectively.
%A fit to only $\phi\ks$ decays yields
%$\cals = \SphiksResultStat$ and $\cala = \AphiksResultStat$.
%A fit to only $\phi\kl$ decays yields
%$-\cals = \mSphiklResultStat$ and $\cala = \AphiklResultStat$.
A $\sinbb$ measurement with DS-II is performed
using $\bz\to\jpsi\ks~(\ks\to \pip\pim~{\rm or}~\piz\piz)$ and
$\bz\to\jpsi\kl$ decays as a crosscheck.
Applying the same procedure to both DS-I and DS-II,
we obtain
$\cals_{\jpsi\kz} = +0.696\pm0.061$(stat) 
and
$\cala_{\jpsi\kz} = +0.011\pm0.043$(stat)
for DS-I, and
$\cals_{\jpsi\kz} = +0.629\pm0.069$(stat) 
and
$\cala_{\jpsi\kz} = +0.035\pm0.044$(stat)
for DS-II.
The results are in good agreement with each other, and
are also consistent with SM expectations.
As all the other checks mentioned above also yield
results consistent with expectations,
we conclude that the difference in $\cals_{\phi\kz}$
between the two datasets is due to a statistical fluctuation.

%%%%%%%%%%%%%%%%%%%%%%
%\section{Discussion}
%%%%%%%%%%%%%%%%%%%%%%

For the $\bz\to\eta'\ks$, the statistical significance of the $CP$ 
asymmetry is calculated with the Feldman-Cousins frequentist
approach~\cite{bib:FC}.
The case with no $CP$ asymmetry ($\cals=0$ and $\cala=0$) is 
ruled out at 99.92\% confidence level, equivalent to 3.4 standard deviations
for Gaussian errors.

\begin{figure}
\resizebox{0.9\textwidth}{!}{\includegraphics{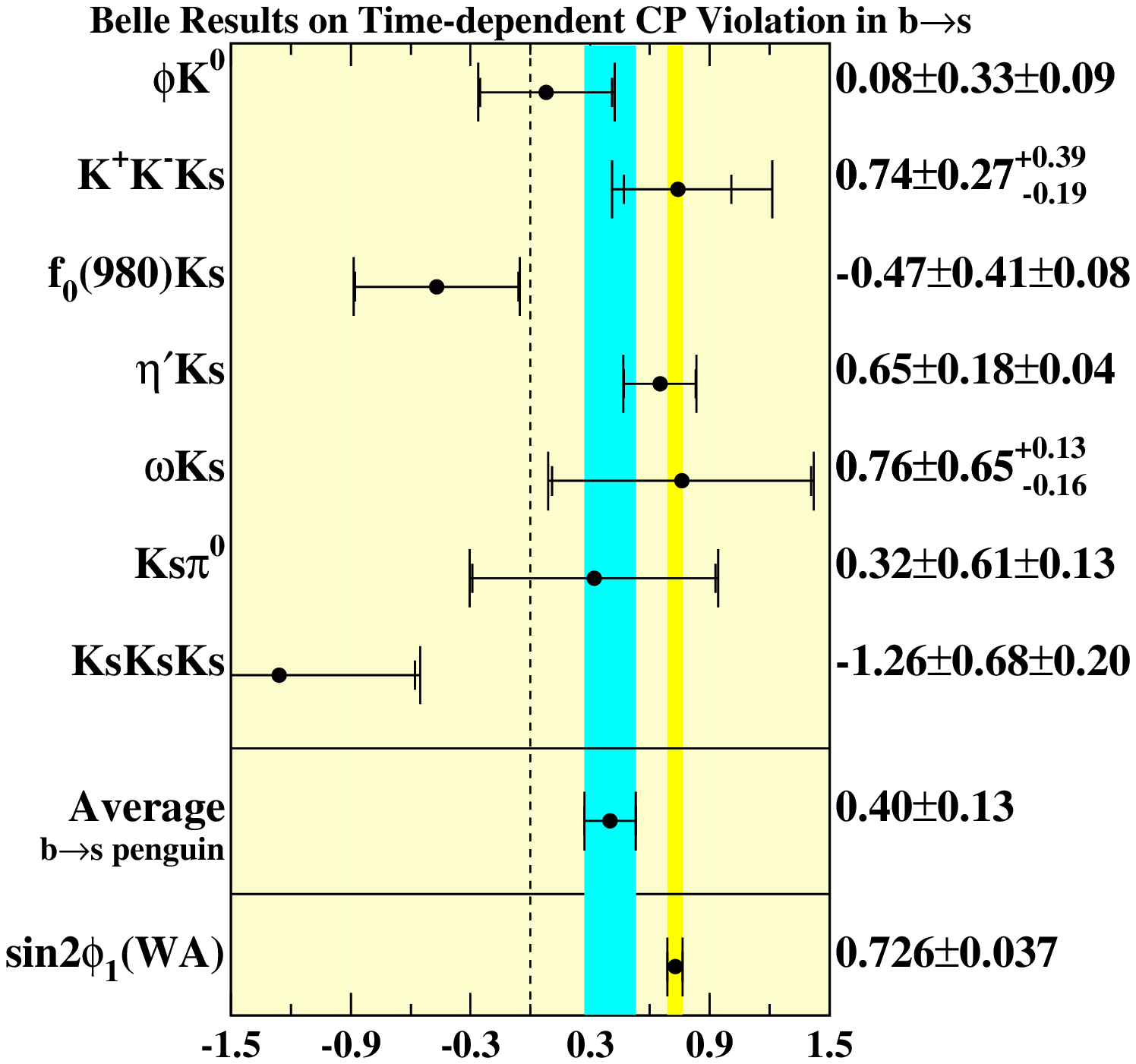}}
\caption{Summary of $\sinbb$ measurements performed with
hadronic $\bz$ decays governed by the $b\to s \overline{q}q$ transition.
The world average $\sinbb$ value obtained from $\bz\to\jpsi\kz$
and other related decay modes governed by the $b\to c\overline{c}s$
transition~\cite{bib:HFAG} is also shown as the SM reference.}
\label{fig:avg}
\end{figure}

As discussed in Section~\ref{sec:introduction},
to a good approximation,
the SM predicts $\cals = -\xi_f\sin 2\phi_1$
%for the $\bz\to\phi\kz$, $\kp\km\ks$, $\fzero\ks$, $\eta'\ks$,
for the $\bz\to\phi\kz$, $\fzero\ks$, $\eta'\ks$,
$\omega\ks$ and $\ks\piz$ decays. 
For the $\bz\to\kp\km\ks$ decay,
the SM prediction is given by $\cals = -(2f_{+}-1)\sin 2\phi_1$.
The effective $\sin 2\phi_1$ for this mode is obtained to be  
$0.74 \pm 0.27 \pm 0.06^{+0.38}_{-0.19}$.
The third error is an additional systematic error arising from the 
uncertainty of the $\xi_f = +1$ fraction.
%an additional systematic error that arises
%from the uncertainty of the $CP$-even component
%fraction ($^{+0.17}_{-0.00}$) is added in quadrature.
Figure~\ref{fig:avg} summarizes 
the $\sinbb$ determination based on our $\cals$ measurements
for these decays.
For each mode, the first error shown in the figure
is statistical and the second error
is systematic. 
We also include the result of the time-dependent $CP$-violating asymmetry 
measurement in $B^0\to \ks\ks\ks$ decays by Belle~\cite{sumisawa:KsKsKs}.
%For the $\bz\to\kp\km\ks$ decay,
%an additional systematic error that arises
%from the uncertainty of the $CP$-even component
%fraction ($^{+0.17}_{-0.00}$) is added in quadrature.
We obtain $\sinbb = \SbsqqResult$ as a weighted average,
where the error includes both statistical and systematic errors.
The result differs from the SM expectation by
2.4 standard deviations.

%%%%%%%%%%%%%%%%%%
\section{Summary}
%%%%%%%%%%%%%%%%%%
We have performed improved measurements of 
$CP$-violation parameters for $\bz \to \phi \kz$ (including
both $\phi\ks$ and $\phi\kl$), $K^+K^-\ks$ and $\eta'\ks$ decays, 
and new measurements for
$\bz\to\fzero\ks$, $\omega\ks$ and $\ks\piz$ decays.
These charmless decays
are dominated by $b\to s$ flavor-changing neutral currents
and are sensitive to possible new $CP$-violating phases.
The results for each individual decay mode are consistent with
the SM expectation within two standard deviations
except for the $\bz\to\fzero\ks$ decay.
The combined result for the $\bz\to\phi\kz$, $\kp\km\ks$, $\fzero\ks$, 
$\eta'\ks$,  $\omega\ks$,  $\ks\piz$ and previously reported $\ks\ks\ks$ 
decays differs from the SM expectation by 2.4 standard deviations.
%The combined result for
%the $\bz\to\phi\kz$, $\kp\km\ks$, $\fzero\ks$, $\eta'\ks$,
%$\omega\ks$ and $\ks\piz$ decays
%differs from the SM expectation by 2.4 standard deviations.
Measurements with a much larger data sample are required to 
conclusively establish the existence of a new $CP$-violating phase
beyond the SM.

%%%%%%%%%%%%%%%%%%%%%%%%%%%
\section*{Acknowledgments}
%%%%%%%%%%%%%%%%%%%%%%%%%%%
%\input{ichep04ack.tex}
%ICHEP04 \include{bnack}
%***** Acknowledgments *****
% use these two starting with pub # 98  Jan 2005
% updated 2/17/05
%----------- Long version, for most papers ----------- 
We thank the KEKB group for the excellent operation of the
accelerator, the KEK cryogenics group for the efficient
operation of the solenoid, and the KEK computer group and
the National Institute of Informatics for valuable computing
and Super-SINET network support. We acknowledge support from
the Ministry of Education, Culture, Sports, Science, and
Technology of Japan and the Japan Society for the Promotion
of Science; the Australian Research Council and the
Australian Department of Education, Science and Training;
the National Science Foundation of China under contract
No.~10175071; the Department of Science and Technology of
India; the BK21 program of the Ministry of Education of
Korea and the CHEP SRC program of the Korea Science and
Engineering Foundation; the Polish State Committee for
Scientific Research under contract No.~2P03B 01324; the
Ministry of Science and Technology of the Russian
Federation; the Ministry of Higher Education, Science and Technology of the
Republic of Slovenia;  the Swiss National Science Foundation; the National
Science Council and
the Ministry of Education of Taiwan; and the U.S.\
Department of Energy.

%%%%%%%%%%%%%%%%%%%%%%%%%%
% Bibliography
%%%%%%%%%%%%%%%%%%%%%%%%%%

\clearpage
\newpage

%%%%%%%%%%%%%%%%%%%%%%%
% Tables and Figures
%%%%%%%%%%%%%%%%%%%%%%%
%%%%%%%%%%%%%%%%%%%%%%%%%%%%%%%%%
\begin{table*}
  \caption{Event fractions $\epsilon_l$,
    wrong-tag fractions $w_l$, wrong-tag fraction differences $\dwl$,
    and average effective tagging efficiencies
    $\eeff^l = \epsilon_l(1-2w_l)^2$ for each $r$ interval for DS-II.
    Errors for $w_l$ and $\dwl$
    include both statistical and systematic uncertainties.
    The event fractions are obtained from $\jpsi\ks$ data.}
  \begin{ruledtabular}
    \begin{tabular}{ccclll}
      $l$ & $r$ interval & $\epsilon_l$ &\multicolumn{1}{c}{$w_l$} 
          & \multicolumn{1}{c}{$\dwl$}  &\multicolumn{1}{c}{$\eeff^l$} \\
      \hline
% ICHEP04      
% 1 & 0.000 -- 0.250 & $0.397\pm 0.015$ & $0.464\pm0.007$ &$+0.010\pm0.007$ &$0.002\pm0.001$ \\
% 2 & 0.250 -- 0.500 & $0.146\pm 0.009$ & $0.321\pm0.008$ &$-0.022\pm0.010$ &$0.019\pm0.002$ \\
% 3 & 0.500 -- 0.625 & $0.108\pm 0.008$ & $0.224\pm0.011$ &$+0.031\pm0.011$ &$0.033\pm0.004$ \\
% 4 & 0.625 -- 0.750 & $0.107\pm 0.008$ & $0.157\pm0.010$ &$+0.002\pm0.011$ &$0.051\pm0.005$ \\
% 5 & 0.750 -- 0.875 & $0.098\pm 0.007$ & $0.109\pm0.009$ &$-0.028\pm0.011$ &$0.060\pm0.005$ \\
% 6 & 0.875 -- 1.000 & $0.144\pm 0.009$ & $0.016\pm0.005$ &$+0.007\pm0.007$ &$0.135\pm0.009$ \\
% BUGFIXED
 1 & 0.000 -- 0.250 & $0.397\pm 0.015$ & $0.464\pm0.007$ &$+0.008\pm0.007$ &$0.002\pm0.001$ \\
 2 & 0.250 -- 0.500 & $0.146\pm 0.009$ & $0.320\pm0.008$ &$-0.022\pm0.010$ &$0.019\pm0.002$ \\
 3 & 0.500 -- 0.625 & $0.108\pm 0.008$ & $0.225\pm0.011$ &$+0.029\pm0.011$ &$0.033\pm0.004$ \\
 4 & 0.625 -- 0.750 & $0.107\pm 0.008$ & $0.158\pm0.010$ &$+0.003\pm0.011$ &$0.050\pm0.005$ \\
 5 & 0.750 -- 0.875 & $0.098\pm 0.007$ & $0.109\pm0.009$ &$-0.028\pm0.011$ &$0.060\pm0.005$ \\
 6 & 0.875 -- 1.000 & $0.144\pm 0.009$ & $0.015\pm0.005$ &$+0.007\pm0.007$ &$0.135\pm0.009$ \\
%total 0.299 +- 0.011 
    \end{tabular}
  \end{ruledtabular}
\label{tab:wtag}
\end{table*}
%%%%%%%%%%%%%%%%%%%%%%%%%%%%%%%%%
%%%%%%%%%%%%%%%%%%%%%%%%%%%%%%%%%%%%%%%%%%%%%%%%
\begin{table}
\caption{
Estimated signal purities and
signal yields $\nsig$ in the signal region for each $\fCP$ mode
that is used to measure $CP$ asymmetries.
The result for the $\bz\to\ks\piz$ decay is obtained
with the sample after flavor tagging but
before vertex reconstruction.
%as
%events that do not have vertex information are also
%used to extract the direct $CP$ violation parameter $\cala$.
The results for other decays are obtained
after flavor tagging and vertex reconstruction.}
\label{tab:num}
\begin{ruledtabular}
\begin{tabular}{llllr}
\multicolumn{1}{c}
{Mode}    &         &$\xi_f$        
                           &\multicolumn{1}{c}{purity} 
                                          & \multicolumn{1}{c}{$\nsig$} \\
\hline
$\phi\ks$         & & $-1$ & $\Pphiks$    & $\Nsigphiks$ \\
$\phi\kl$         & & $+1$ & $\Pphikl$    & $\Nsigphikl$ \\
%$K^+K^-\ks$       & & $+1$ & $\Pkpkmks$   & $\Nsigkpkmks$ \\
$K^+K^-\ks$       & & $+1$ (83\%), $-1$ (17\%) & $\Pkpkmks$   & $\Nsigkpkmks$ \\
$\fzero\ks$       & & $+1$ & $\Pfzeroks$  & $\Nsigfzeroks$ \\
$\eta'\ks$        & & $-1$ & $\Petapks$   & $\Nsigetapks$ \\
$\omega\ks$       & & $-1$ & $\Pomegaks$  & $\Nsigomegaks$ \\
$\ks\piz$ &(high-$\rsigbkg$)& $-1$ & $\PkspizH$   & $\NsigkspizH$ \\
          &(low-$\rsigbkg$) & $-1$ & $\PkspizL$   & $\NsigkspizL$ \\
\end{tabular}
\end{ruledtabular}
\end{table}
%%%%%%%%%%%%%%%%%%%%%%%%%%%%%%%%%%%%%%%%%%%%%%%%
%%%%%%%%%%%%%
\begin{table}
\caption{Results of the fits to the $\Dt$ distributions.
The first error is statistical and the second
error is systematic. We combine
$\bz\to\phi\ks$ and $\bz\to\phi\kl$ decays
to obtain $\cals_{\phi\kz}$ and $\cala_{\phi\kz}$.
}
\label{tab:result}
\begin{ruledtabular}
\begin{tabular}{lcll}
\multicolumn{1}{c}{Mode} &  
SM expectation for $\cals$ &
\multicolumn{1}{c}{$\cals$} & 
\multicolumn{1}{c}{$\cala$} \\
\hline
$\phi\kz$   & $+\sinbb$   & $\SphikzResult$    & $\AphikzResult$    \\
$K^+K^-\ks$ & $-(2f_{+}-1)\sinbb$   & $\SkpkmksResult$   & $\AkpkmksResult$   \\
$\fzero\ks$ & $-\sinbb$   & $\SfzeroksResult$  & $\AfzeroksResult$  \\
$\eta'\ks$  & $+\sinbb$   & $\SetapksResult$   & $\AetapksResult$   \\
$\omega\ks$ & $+\sinbb$   & $\SomegaksResult$  & $\AomegaksResult$  \\
$\ks\piz$   & $+\sinbb$   & $\SkspizResult$    & $\AkspizResult$    \\
\end{tabular}
\end{ruledtabular}
\end{table}
%%%%%%%%%%%
%%%%%%%%%%%%%%%%%%%%%%%%%%%%%%%%%
\begin{table*}
  \caption{Summary of the systematic errors on $\cals$.}
  \begin{ruledtabular}
    \begin{tabular}{lrrrrrr}
           & $\phi\kz$ 
                    &$\kp\km\ks$ 
                           &$\fzero\ks$ 
                                   &$\eta'\ks$
                                            &$\omega\ks$
                                                              &$\ks\piz$\\
\hline
Vertex 
reconstruction & 0.01 & 0.01  & 0.02  & 0.01  & 0.01   & 0.02 \\
Flavor tagging & 0.01 &$<0.01$& 0.01  & 0.01  & 0.04   & 0.01 \\
Resolution 
function       & 0.04 & 0.03  & 0.03  & 0.03  & 0.07   & 0.05 \\
Physics 
parameters     & $<0.01$ &$<0.01$& 0.01  &$<0.01$& 0.01   & 0.02 \\
Possible 
fit bias       & 0.01 & 0.01  & 0.03  & 0.01&$^{+0.01}
                                              _{-0.10}$& 0.02 \\
Background 
fraction  & 0.08 & 0.02  & 0.05  & 0.02  & 0.10   & 0.08 \\
Background 
$\Dt$ shape    & 0.01 &$<0.01$& 0.04  &$<0.01$& 0.02   & 0.08 \\
Tag-side 
interference& $<0.01$ &$<0.01$&$<0.01$&$<0.01$& 0.01   &$<0.01$\\
\hline      
Total          &$\SphikzSyst$ 
                      &$\SkpkmksSyst$
                              &$\SfzeroksSyst$
                                      &$\SetapksSyst$
                                              &$\SomegaksSyst$
                                                               &$\SkspizSyst$\\
    \end{tabular}
  \end{ruledtabular}
\label{tab:ssyserr} 
\end{table*}
%%%%%%%%%%%%%%%%%%%%%%%%%%%%%%%%%
%%%%%%%%%%%%%%%%%%%%%%%%%%%%%%%%%
\begin{table*}
  \caption{Summary of the systematic errors on $\cala$.}
  \begin{ruledtabular}
    \begin{tabular}{lrrrrrr}
           & $\phi\kz$ 
                      &$\kp\km\ks$ 
                               &$\fzero\ks$ 
                                       &$\eta'\ks$
                                                &$\omega\ks$
                                                          &$\ks\piz$\\
\hline
Vertex 
reconstruction & 0.03   & 0.04   & 0.04   & 0.03   & 0.04   & 0.04 \\
Flavor tagging &$<0.01$ &$<0.01$ & 0.01   & 0.01   &$<0.01$ & 0.01 \\
Resolution 
function       & 0.02   & 0.02   & 0.02   & 0.01   & 0.04   &$<0.01$\\
Physics 
parameters     &$<0.01$ &$<0.01$ &$<0.01$ &$<0.01$ &$<0.01$ &$<0.01$\\
Possible 
fit bias       & 0.01   & 0.01   & 0.03   & 0.01 &$^{+0.01}
                                                   _{-0.03}$& 0.01 \\
Background 
fraction       & 0.04   & 0.01   & 0.06   & 0.02   & 0.14   & 0.04 \\
Background 
$\Dt$ shape    & 0.03   &$<0.01$ & 0.01   &$<0.01$ & 0.03   & 0.05 \\
Tag-side 
interference   & 0.06   & 0.06   & 0.04   & 0.03   & 0.03   & 0.05\\
\hline      
Total          &$\AphikzSyst$ 
                        &$\AkpkmksSyst$
                                &$\AfzeroksSyst$
                                          &$\AetapksSyst$
                                                   &$\AomegaksSyst$
                                                            &$\AkspizSyst$\\
    \end{tabular}
  \end{ruledtabular}
\label{tab:asyserr} 
\end{table*}
%%%%%%%%%%%%%%%%%%%%%%%%%%%%%%%%%

\clearpage
\newpage
%%%%%%%%%%%%%%%%%%%%%%%%%%%%%%%%%%%%%%%%%%%%%%%%
%%%%%%%%%%%%%%%%%%%%%%%%%%%%%%%%%%%%%%%%%%%%%%%%
\clearpage
\newpage
%%%%%%%%%%%%%%%%%%%%%%%%%%%%%%%%%%%%%%%%%%%%%%%%
%%%%%%%%%%%%%%%%%%%%%%%%%%%%%%%%%%%%%%%%%%%%%%%%
\clearpage
\newpage
%%%%%%%%%%%%%%%%%%%%%%%%%%%%%%%%%%%%%%%%%%%%%%%%
%%%%%%%%%%%%%%%%%%%%%%%%%%%%%%%%%%%%%%%%%%%%%%%%
\clearpage
\newpage
%%%%%%%%%%%%%%%%%%%%%%%%%%%%%%%%%%%%%%%%%%%%%%%%%%%%%%%
%%%%%%%%%%%%%%%%%%%%%%%%%%%%%%%%%%%%%%%%%%%%%%%%%%%%%%%
\clearpage
\newpage

%%%%%%%%%%%%%%%%%%%%%%%%%%%%%%%%%%%%%%%%%%%%%%%%%%%%%%%
%%%%%%%%%%%%%%%%%%%%%%%%%%%%%%%%%%%%%%%%%%%%%%%%%%%%%%%


\begin{thebibliography}{999}
\bibitem{bib:CC}
Throughout this paper,
the inclusion of the charge conjugate decay mode is implied
unless otherwise stated.
\bibitem{footnote:phikz}
We use an alternative definition of
the raw asymmetry, 
$(N_{q\xi_f = -1}-N_{q\xi_f = +1})/(N_{q\xi_f = -1}+N_{q\xi_f = +1})$, 
for the decay $\bz\to\phi\kz$ to take the opposite $CP$ parities
for the decays $\bz\to\phi\ks$ and $\phi\kl$ into account.
\end{thebibliography}
\end{document}